\documentclass[twocolumn,times]{aastex62}  
\usepackage{graphicx}
\usepackage{float}
\usepackage{color}
\usepackage{soul}
\usepackage{amsmath}
\usepackage{xspace}
\usepackage{makecell}

\begin{document}

\title{Determining the Structure of Rotating Massive Stellar Cores with Gravitational Waves}

\author[0000-0002-4983-4589]{Michael A. Pajkos}
\affiliation{Department of Physics and Astronomy, Michigan State University, East Lansing, MI 48824, USA}
\affiliation{Department of Computational Mathematics, Science, and Engineering, Michigan State University, East Lansing, MI 48824, USA}
\affiliation{Joint Institute for Nuclear Astrophysics-Center for the Evolution of the Elements, Michigan State University, East Lansing, MI 48824, USA}

\author[0000-0001-9440-6017]{MacKenzie L. Warren}
\affiliation{Department of Physics, North Carolina State University, Raleigh, NC 27695, USA}
\affiliation{Joint Institute for Nuclear Astrophysics-Center for the Evolution of the Elements, Michigan State University, East Lansing, MI 48824, USA}
\affiliation{Department of Physics and Astronomy, Michigan State University, East Lansing, MI 48824, USA}
\affiliation{National Science Foundation Astronomy and Astrophysics Fellow}

\author[0000-0002-5080-5996]{Sean M.~Couch}
\affiliation{Department of Physics and Astronomy, Michigan State University, East Lansing, MI 48824, USA}
\affiliation{Department of Computational Mathematics, Science, and Engineering, Michigan State University, East Lansing, MI 48824, USA}
\affiliation{Joint Institute for Nuclear Astrophysics-Center for the Evolution of the Elements, Michigan State University, East Lansing, MI 48824, USA}
\affiliation{National Superconducting Cyclotron Laboratory, Michigan State University, East Lansing, MI 48824, USA}


\author[0000-0002-8228-796X]{Evan P. O'Connor}
\affiliation{Department of Astronomy and The Oskar Klein Centre, Stockholm University, AlbaNova, SE-106 91 Stockholm, Sweden}

\author[0000-0002-1473-9880]{Kuo-Chuan Pan}
\affiliation{Department of Physics, National Tsing Hua University, Hsinchu 30013, Taiwan}
\affiliation{Institute of Astronomy, National Tsing Hua University, Hsinchu 30013, Taiwan}
\affiliation{Center for Informatics and Computation, National Tsing Hua University, Hsinchu 30013, Taiwan}
\affiliation{National Center for Theoretical Sciences, National Tsing Hua University, Hsinchu 30013, Taiwan}

\published{17 June 2021 in The Astrophysical Journal}


\shorttitle{Constraining Progenitor Properties with GWs}
\shortauthors{Pajkos et al.}

 \begin{abstract}
  
  The gravitational wave (GW) signal resulting from stellar core collapse encodes a wealth of information about the physical parameters of the progenitor star and the resulting core-collapse supernova (CCSN). 
  We present a novel approach to constrain CCSN progenitor properties at collapse using two of the most detectable parts of the GW signal: the core-bounce signal and evolution of the dominant frequency mode from the protoneutron star.  
  We focus on the period after core bounce but before explosion and investigate the predictive power of GWs from rotating CCSNe to constrain properties of the progenitor star.  
  We analyze 34 2D and four 3D neutrino-radiation-hydrodynamic simulations of stellar core collapse in progenitors of varied initial mass and rotation rate.
  Extending previous work, we verify the compactness of the progenitor at collapse to correlate with the early ramp-up slope, and in rotating cases, also with the core angular momentum.  
  Combining this information with the bounce signal, we present a new analysis method to constrain the pre-collapse core compactness of the progenitor.  
  Because these GW features occur less than a second after core bounce, this analysis could allow astronomers to predict electromagnetic properties of a resulting CCSN even before shock breakout.
  
   \keywords{core-collapse supernovae (304) -- gravitational wave astronomy (675) -- gravitational wave sources (677) -- massive stars (732)}
\end{abstract}

%

\section{Introduction}

Gravitational waves (GWs) provide astronomers with an entirely new spectrum of signals to detect, coming from a variety of astrophysical processes.  As current GW observatories---Advanced Laser Interferometer Gravitational-wave Observatory (aLIGO), Advanced Virgo, and Kamioka Gravitational Wave Detector (KAGRA)---continue observing runs and with future GW observatories on the horizon---Deci-hertz
Interferometer Gravitational Wave Observatory (DECIGO), Einstein Telescope, LIGO-India, and the Laser Interferometer Space Antenna (LISA)---the number of GW detections will only increase \citep{gossan:2016}.  One site of particular interest for some of these observatories is the explosive endings of massive stars: core-collapse supernovae (CCSNe).  CCSNe are important in a broader astrophysical context because they contain matter with densities over many orders of magnitude.  Acting as unique laboratories, better understanding these stellar explosions has a broad impact on many areas of astronomy: predicting compact object birth, restricting the nuclear equation of state (EOS), and constraining stellar rotation, to name a few.

After being produced in the center of a CCSN, GWs pass through the outer stellar envelope unobstructed.  For decades, astronomers have attempted to leverage this unique characteristic to prepare for the next CCSN event by connecting the features of a GW signal with the internal physics of the supernova or protoneutron star (PNS) inside.  Currently, the detection range for GWs from CCSNe is Galactic in scale, leaving the expected rate to be $\sim 2$ events per century \citep{diehl:2006}.  Given the rarity of potential observations, GW predictions from numerical models have been important to prepare GW astronomers for the next CCSN event. 

While the CCSN problem involves a variety of physics, a proper treatment of gravity is one of the most important aspects when predicting GWs.  While a numerical scheme that simulates a dynamically evolving space-time \citep[e.g.][]{shibata:1995,baumgarte:1999} may be ideal for accuracy, the immense computational cost can prevent incorporating other robust physics features, such as magnetic fields or neutrinos.  Approximations have been made to capture some of the general relativistic (GR) features for a given mass distribution within a supernova.  Previous works have paired Newtonian hydrodynamics with an effective GR gravitational potential (GREP) \citep{rampp:2002,marek:2006,bruenn:2016,moro:2018,oconnor:2018}.  Consistent with a modified Tolman-Oppenheimer-Volkhoff equation, GREP empirically satisfies the solution to hydrostatic equilibrium and has been shown to fairly accurately reproduce overall features of CCSN numerical models \citep{rampp:2002,marek:2006,muller:2012,oconnor:2018,schneider:2020}.  Another, more advanced, approximation is the so-called conformal flatness condition (CFC), in which the spatial three metric is approximated by the flat space-time three metric.  This scheme qualitatively agrees with CCSN results using GREP \citep{shibata:2004} and reproduces early CCSN GW signals within a few percent of similar simulations that directly solve Einstein's field equations \citep{ott:2007}.  As a variant, CFC has also been reformulated as an \textit{augmented CFC} scheme to address uniqueness issues of the elliptic constraints present in CFC \citep{saijo:2004,cordero-carrion:2009,bmuller:2019}. 

Once the core temperature for a massive star ($\gtrsim 8 \,M_\odot$) becomes sufficiently high ($T \sim 5 $ GK), iron nuclei begin to photodissociate and undergo electron capture.  These processes result in a net loss of pressure, triggering the core to collapse inward.  At sufficiently high densities ($\sim 2\times 10^{14}\, \textrm{g\,cm}^{-3}$), the nuclear force halts the matter infall on the timescale of microseconds and, in the case of rotating CCSNe, a burst of GWs is produced: the GW \textit{bounce signal}.  The bounce signal has been studied extensively and it is found that---except in extreme scenarios---more angular momentum in the supernova center will produce a bounce signal with higher amplitude \citep{muller:1982,moench:1991,yamada:1995,zwerger:1997,dimm:2002,kotake:2003,shibata:2004,abdik:2014}.

As the subsonic inner core meets the supersonic outer core, a shock front ensues, photodissociating material at larger radii as it propagates outward.  This process forms a negative lepton gradient via neutrino production, causing prompt convection in the post-shock region \citep{mazurek:1982,bruenn:1985,bruenn:1989,burrows:1992}.  This prompt convection is an important feature in CCSN evolution and has been shown to directly contribute to the GW signal \citep{marek:2009b,muller:2017,richers:2017,nagakura:2018}.

As the shock front continues to propagate outward, the matter motion behind the shock is subject to a variety of instabilities that can emit GW signals: post-shock convection \citep{burrows:1996,muller:1997,muller:2004,murphy:2009,muller:2013}, the standing accretion shock instability (SASI) \citep{blondin:2003,blondin:2006,ohnishi:2006,foglizzo:2007,scheck:2008,iwakami:2009,fernandez:2010,kuroda:2016,andresen:2017}, and the PNS vibrational modes \citep{muller:1997,cerda-duran:2013,torres-forne:2018,torres-forne:2019}. 

Perturbation theory has been used historically to provide analytic estimates of PNS properties by identifying the resonant frequencies (GW astroseismology), or vibrational modes---for example, g-, f-, p-, r-, and w-modes \citep{unno:1989}.  As outlined by \citet{gautschy:1995}, the restoring force for p-modes comes from the pressure of the gas.  The restoring force for g-modes is the buoyancy force.  \citet{andersson:1998} and \citet{kokkotas:1999} describe r-modes arising in rotating stars and grow unstable due to the emission of GWs from the stellar interior.  \citet{kokkotas:1992} identify the presence of so-called w-modes that are closely related to the oscillations in the space-time metric.
\citet{andersson:1996} built on this work by suggesting the f-mode evolves with the average density of the star and the damping rate of the w-mode depends linearly on compactness.  

More recently, \citet{sotani:2016} used multiple 1D simulations to show that PNS oscillation frequencies are almost independent of PNS electron fraction ($Y_e$) and entropy per baryon profiles. \citet{sotani:2017} use 3D models to relate the w$_1$-mode to the PNS mass and radius.  \citet{sotani:2019} use 1D simulations to examine PNS structure during the accretion phase, en route to black hole (BH) formation; moreover, various groups have even used GWs to probe BH formation itself for a variety of progenitor masses \citep{ott:2011,cerda-duran:2013,kuroda:2018,pan:2018}.  Other works investigate the influence of rotation and magnetic fields in the core-collapse scenario \citep{obergaulinger:2017,obergaulinger:2018}.  \citet{warren:2020} recently explored the GW signal a few seconds after bounce, displaying correlations between initial progenitor compactness and the slope of the GW frequency emitted from the dominant PNS mode, in frequency versus time space (hereafter referred to as ``ramp-up slope'').  \citet{sotani:2020} explore the dimensional dependence of GW generation from the PNS.  They point out correlations between the relative strengths of different modes compared to PNS characteristics: average density and compactness.  Likewise, research continues into how these different modes interact via \textit{avoided crossing} \citep{sotani:2020b}.   \citet{moro:2018} examine GW emission for a moderately rotating CCSN ($\Omega_{\textrm{central}} \sim 0.2$ rad s$^{-1}$).  \citet{radice:2019} and \citet{oconnor:2018b} use a suite of 3D simulations to show how turbulent kinetic energy accreted by the PNS relates to the GW energy radiated.  \citet{mezzacapa:2020} recently explored the GW production, by region, from a $15\, M_\odot$ star.   \citet{vartanyan:2019} relate multimessenger signals from CCSNe to physical properties at the center of the supernova.  Like previous works, \citet{vartanyan:2020} investigate GW production from neutrino emission asymmetries in CCSNe \citep{muller:1997,kotake:2009}.  \citet{pan:2020} and \citet{shibagaki:2021} study GW emission from rotating 3D progenitors.  And in even more exotic scenarios, \citet{zha:2020} identify the GW signals expected from a quantum chromodynamic phase transition of a protocompact star, originating from a CCSN.
 
  As discussed, many studies that use multidimensional simulations have either explored only the bounce signal, in the case of rotation, or the accretion phase signal from nonrotating supernovae.  As all stars rotate to some degree, an opportunity arises to investigate the effect of rotation on GWs emitted from CCSNe during the accretion phase.  Furthermore, there exists a growing need in the supernova community to not only predict gravitational waveforms but extract information from them in new ways, in order to constrain physical properties of the progenitor star.  In this work, we show the GW signal from the dominant PNS mode encodes angular momentum information of the CCSN at bounce in a quantifiable way.  We also present a novel analysis that combines multiple features of the GW signal from a single rotating CCSN to help constrain the properties of the progenitor star.  
  
  The strength of this technique stems from its observational considerations.  Previous works depend on tracking multiple, relatively weaker modes of a PNS to constrain mass and radius.  This work only depends on the loudest components of the GW signal and are thus most likely to be reconstructed by current GW detectors \citep{mciver:2015}.  Furthermore, this analysis is valuable because it uses multimessenger signals that are emitted less than a second after core bounce to constrain core compactness.  Applying previous works that draw correlations between core compactness and electromagnetic (EM) observables days after a supernova explosion \citep{sukhbold:2016}, our work provides a predictive framework that would allow astronomers to anticipate EM properties of the supernova, even before shock breakout occurs.


This paper is organized as follows:  in Section \ref{sec:method} we present our methods and treatment of microphysics within our \texttt{FLASH} simulations.  Section \ref{sec:results} contains our analysis and outlines observational considerations.  Finally, in Section \ref{sec:summary} we discuss and conclude. 

\begin{table*}[t]
\centering
\begin{tabular}{c|c|c|c|c|c|c|c|c|c|c|c}
 Label & M($M_\odot$)  & $\Omega_0$(rad s$^{-1}$) & \textit{A}($10^3$ km) & EOS & 2D/3D & $\nu$ Treatment & $e^-$ rates & $M_\mathrm{core}^\mathrm{B} (M_\odot)$ & $Y^\mathrm{c}_e$ & $\beta^\mathrm{B}_\mathrm{core}$& $t^\mathrm{pb}_\mathrm{end}$ (s)  \\
\hline
s12o0 & 12  & 0  &  0.8123 & SFHo & 2D & M1 & SNA & 0.578 & 0.278 & 0.000 & 0.3    \\
s12o0.5 & 12  & 0.5  &  0.8123 & SFHo & 2D & M1 & SNA & 0.580 & 0.279 & 0.001 & 0.3   \\
s12o1 & 12  & 1  &  0.8123 & SFHo & 2D & M1 & SNA & 0.581 & 0.279 & 0.006 &0.3   \\
s12o2 & 12  & 2  &  0.8123 & SFHo & 2D & M1 & SNA & 0.584 & 0.279 & 0.021 &0.3   \\
s12o3 & 12  & 3  &  0.8123 & SFHo & 2D & M1 &
SNA & 0.576 & 0.280 & 0.042 &0.3   \\
s20o0 & 20  & 0  &  1.021 & SFHo & 2D & M1 & SNA & 0.564 & 0.273 & 0.000 & 0.3   \\
s20o.5 & 20  & 0.5  &  1.021 & SFHo & 2D & M1 & SNA & 0.567 & 0.273 & 0.006 &0.3   \\
s20o1 & 20  & 1  &  1.021 & SFHo & 2D & M1 & SNA & 0.572 & 0.274 & 0.021 &0.3   \\
s20o2 & 20  & 2  &  1.021 & SFHo & 2D & M1 & SNA & 0.568 & 0.274 & 0.066 &0.3   \\
s20o3 & 20  & 3  &  1.021 & SFHo & 2D & M1 & SNA & 0.534 & 0.274 & 0.106 &0.3   \\
s40o0 & 40  & 0  & 1.282  & SFHo & 2D & M1 & SNA & 0.556 & 0.267 & 0.000 &0.3   \\
s40o0.5 & 40  & 0.5  & 1.282  & SFHo & 2D & M1 & SNA & 0.560 & 0.268 & 0.011 &0.3   \\
s40o1 & 40  & 1  & 1.282  & SFHo & 2D & M1 & SNA & 0.562 & 0.268 & 0.037 &0.3   \\
s40o2 & 40  & 2  & 1.282  & SFHo & 2D & M1 & SNA & 0.554 & 0.268 & 0.101 &0.3   \\
s60o0 & 60  & 0  & 0.9112  & SFHo & 2D & M1 & SNA & 0.571 & 0.276 & 0.000 &0.3   \\
s60o0.5& 60  & 0.5  & 0.9112  & SFHo & 2D & M1 & SNA & 0.574 & 0.276 & 0.003 &0.3   \\
s60o1 & 60  & 1  & 0.9112  & SFHo & 2D & M1 & SNA & 0.576 & 0.277 & 0.013 &0.3   \\
s60o2 & 60  & 2  & 0.9112  & SFHo & 2D & M1 & SNA & 0.571 & 0.277 & 0.043 &0.3   \\
s60o3 & 60  & 3  & 0.9112  & SFHo & 2D & M1 & SNA & 0.557 & 0.278 & 0.076 &0.3   \\ \hline
s12o0x$^\ddagger$ & 12  & 0  &  0.8123 & SFHx & 2D & M1 & SNA & 0.563 & 0.273 & 0.000 & 0.3   \\
s12o0.5x & 12  & 0.5  &  0.8123 & SFHx & 2D & M1 & SNA & 0.564 & 0.273 & 0.001 &0.005   \\
s12o2x$^\ddagger$ & 12  & 2  &  0.8123 & SFHx & 2D & M1 & SNA & 0.580 & 0.273 & 0.022 & 0.3   \\
s12o3x$^\ddagger$ & 12  & 3  &  0.8123 & SFHx & 2D & M1 & SNA & 0.578 & 0.272 & 0.042 &0.3   \\
s20o0.5x & 20  & 0.5  &  1.021 & SFHx & 2D & M1 & SNA & 0.550 & 0.267 & 0.006 &0.005   \\
s20o1x & 20  & 1  &  1.021 & SFHx & 2D & M1 & SNA & 0.568 & 0.268 & 0.022 &0.005   \\
s20o2x & 20  & 2  &  1.021 & SFHx & 2D & M1 & SNA & 0.570 & 0.267 & 0.068 & 0.005   \\
s20o3x & 20  & 3  &  1.021 & SFHx & 2D & M1 & SNA & 0.507 & 0.270 & 0.103 &0.005   \\
s40o0.5x & 40  & 0.5  & 1.282  & SFHx & 2D & M1 & SNA & 0.562 & 0.262 & 0.011 &0.005   \\
s60o0.5x & 60  & 0.5  & 0.9112  & SFHx & 2D & M1 & SNA & 0.557 & 0.271 &0.003 &0.005   \\
s60o1x & 60  & 1  & 0.9112  & SFHx & 2D & M1 & SNA & 0.558 & 0.271 & 0.012 &0.005   \\
s60o2x & 60  & 2  & 0.9112  & SFHx & 2D & M1 & SNA & 0.572 & 0.271 & 0.044 &0.005   \\
s60o3x & 60  & 3  & 0.9112  & SFHx & 2D & M1 & SNA & 0.562 & 0.271 & 0.078 &0.005   \\ \hline
s12o2$^\nu$ & 12  & 2  &  0.8123 & SFHo & 2D & M1 & LMP+N50 & 0.520 & 0.259 & 0.020 &0.005   \\
s60o2$^\nu$ & 60  & 2  & 0.9112  & SFHo & 2D & M1 & LMP+N50 & 0.535 & 0.259 & 0.044 &0.005  \\ \hline
s27o2$^{3D}$ & 27  & 2  & 0.7700  & LS220 & 3D & M1$^*$ & SNA & N/A & N/A & N/A & 0.005  \\
s40o0$^{3D}$ & 40  & 0  & 1  & LS220 & 3D & Ye($\rho$)+IDSA & SNA & 0.576 & 0.272 &0.000 & 0.3  \\
s40o0.5$^{3D}$ & 40  & 0.5  & 1  & LS220 & 3D & Ye($\rho$)+IDSA & SNA & 0.509 & 0.272 & 0.008 & 0.3  \\
s40o1$^{3D}$ & 40  & 1  & 1  & LS220 & 3D & Ye($\rho$)+IDSA & SNA & 0.562 & 0.272 & 0.028 & 0.3  \\
\end{tabular}
\caption{ Information regarding setup information for all 38 models in this study.  Column labels represent the following: M--zero age main sequence mass, $\Omega_0$--central rotation rate at collapse, $A$ differential rotation parameter, EOS, 2D/3D--dimensionality of the simulation, neutrino ($\nu$) treatment, $e^-$ rates--electron capture rates used to construct neutrino opacity tables, $M_\mathrm{core}^B$--mass of core at bounce, $Y_e^\mathrm{c}$--central $Y_e$ at bounce, $\beta^\mathrm{B}_\mathrm{core}$ is the ratio of rotational kinetic energy to gravitational binding energy at bounce, $t^\mathrm{pb}_\mathrm{end}$--simulation end time (post bounce).  For $e^-$ rates, SNA represents the single nucleus approximation used by \citet{bruenn:1985}.  LMP+N50 represents the Laganke-Martinez Pinedo rates \citep{langanke:2001} supplemented by the calculations of \citet{titus:2018}.  $\ddagger$ denote the three models saved as \textit{test data} for our multidimensional fit in Figure \ref{fig:3D_param} and Equation (\ref{eq:xi_plane}).  For s27o2$^{3D}$ M1$^*$ indicates M1 neutrino transport, without inelastic scattering or velocity dependent terms.}

\label{table:everyone}
\end{table*}


\section{Methods}
\label{sec:method}

In this work, we simulate the core collapse of the 12.0, 20.0, 40, and 60 $M_\odot$ nonrotating, solar-metallicity progenitors models from \citet{sukhbold:2016}.  
We use the \texttt{FLASH} (version 4) multiscale, multiphysics adaptive mesh refinement simulation framework \citep{fryxell:2000,dubey:2009}.\footnote[7]{\url{http://flash.uchicago.edu/site/}}  
Our grid setup is a 2D cylindrical geometry with the PARAMESH (v.4-dev) library for adaptive mesh refinement  \citep{macneice:2000}.  The outer boundary is $10^4$ km in all directions, with nine levels of refinement---a finest grid spacing of about 0.65 km.
The maximum allowed level of refinement is decreased as a function of spherical radius, $r$, in order to maintain a resolution aspect ratio, $\Delta x_i / r$, of about 0.01, corresponding approximately to an \textit{angular} resolution of $0.5^{\circ}$.  We use the GREP for our gravitational treatment \citep{marek:2006, oconnor:2018} used alongside the multipole Poisson solver of \citet{couch:2013a}, where we retain spherical harmonic orders up through 16. 


To model the transport of neutrinos, we incorporate an M1 scheme: a multidimensional, multispecies, energy-dependent, two-moment scheme with an analytic closure.  Our implementation is based on \citet{oconnor:2015}, \citet{shibata:2011}, and \citet{cardall:2013}.  For a detailed outline of the M1 implementation in \texttt{FLASH}, we direct the reader to \citet{oconnor:2018}.
 We use 12 energy bins spaced logarithmically up to 250 MeV.  The full set of rates and opacities we use is described in \citet{oconnor:2017a}.  As outlined by \citet{horowitz:2017}, we use the effective, many-body, corrected rates for neutrino-nucleon, neutral current scattering.  In this study---unlike our previous work \citep{pajkos:2019}---we incorporate velocity-dependent neutrino transport and account for inelastic neutrino-electron scattering.

In total, we simulate 34 CCSNe; we use the SFHo EOS for 19 and the SFHx EOS for three of our simulations that run to 300 ms pb \citep{steiner:2013,steiner:2013b}.  We run additional simulations that run through core bounce, 10 of which use the SFHx EOS and two of which use the SFHo EOS with modified electron capture rates \citep{langanke:2003,steiner:2013,steiner:2013b,sullivan:2016,titus:2018}.  Additionally, we incorporate four 3D simulations into our analysis.  We extract the GW bounce signal from the collapse of one rotating 27 $M_\odot$ progenitor \citep{woosley:2002} that uses the LS220 EOS \citep{lattimer:1991}, M1 neutrino transport, and has a central rotation rate of 2 rad s$^{-1}$.  We examine the bounce signal and accretion phase signal of three simulations that model a 40 $M_\odot$ \citep{woosley:2007} collapse using the LS220 EOS and use the Isotropic Diffusion Source Approximation (IDSA) neutrino treatment \citep{liebendorfer:2009,pan:2016,pan:2018,pan:2020}.  The three central rotational velocities are 0, 0.5, and 1 rad s$^{-1}$.  For a detailed outline of all simulation parameters, see Table \ref{table:everyone}.

\subsection{Rotational Profiles}

To begin, the nonrotating 1D progenitor models are mapped onto our 2D Eulerian grid.  We then apply an artificial rotation profile
\begin{equation}
    \Omega(r) = \Omega_0 \bigg[1 + \bigg(\frac{r}{A}\bigg)^2 \bigg]^{-1}, 
    \label{eq:omega}
\end{equation}
where $r = \sqrt{R^2 + z^2}$ is the spherical radius for a given cylindrical radius $R$ and altitude $z$, $\Omega_0$ is the central angular speed of the star, and $A$ is the differential rotation parameter \citep{eriguchi:1985}.  
Small $A$ values imply a greater degree of differential rotation, while larger $A$ values push the rotation profile closer to solid body. 
By multiplying the angular speed with the distance from the rotation axis, the linear rotational velocity is calculated: $v_\phi (R, z) = R \Omega (r) $. 

The internal rotation rates and profiles of massive stellar cores at collapse are still poorly constrained.  Other work \citep[e.g.,][]{abdik:2014} explore varying the differential rotation parameter $A$ and investigate its impact on the GW bounce signal.
We assign $A$ values based on {\it compactness} \citep{oconnor:2011} of the core at collapse. The core compactness as introduced by \citet{oconnor:2011} is defined as
\begin{equation}\label{eqn:xi}
    \xi_M = \left.\frac{M/M_{\odot}}{R(M_\mathrm{bary}=M)/1000\, \text{km}}\right\vert_\mathrm{collapse} ,
\end{equation} 
where $M$ is the baryonic mass, and $R(M) $ is the radius at the corresponding mass coordinate.  How $A$ relates to $\xi_M$ is based on an empirical fit determined in our previous work \citep{pajkos:2019}.  In short, this relationship quantifies how the rotational velocity assigned at collapse tracks the progenitor core structure based on the models of \citet{heger:2005}.

For our 2D simulations, we select $\Omega_0 = 0, 0.5, 1, 2,$ and $3$ rad s$^{-1}$ for central rotation rates.  \citet{pajkos:2019} showed that the ramp-up slope decreases as the PNS becomes more centrifugally supported for these integer valued rotation rates.  In aims to quantify this relationship we select identical rates.  Due to magnetic breaking, the presence of rapidly rotating stars---similar to stars with $\Omega_0 \geq 1$ rad s$^{-1}$---is quite rare \citep{woosley:2006}.  In an attempt to more finely sample the lower rotation rate parameter space, we extend this previous study by including simulations with $\Omega_0 = 0.5$ rad s$^{-1}$ as well.

To maintain the fidelity of our simulation suite, we choose to omit the 40 $M_\odot$ progenitor at $\Omega_0 = 3$ rad s$^{-1}$ from our following analysis.  The $\xi_{2.5}$ value of this progenitor is nearly double that of the 20 $M_\odot$ progenitor (the next closest compactness value), resulting in nearly solid-body core rotation.   This rotation profile for the 40 $M_\odot$ results in vast amounts of angular momentum, ultimately leading to likely unphysical rotation dominated dynamics, and in particular, a highly suppressed core-bounce GW signal. 

%

\subsection{GW Signal Extraction}

To extract the GW signal from our simulations, we adopt the dominant, quadrupole moment formula for the gravitational strain, through the slow motion, weak-field formalism 
\citep[eg.,][]{blanchet:1990,finn:1990}
\begin{equation}
    h_+ \approx \frac{3}{2}\frac{G}{Dc^4}
    \frac{d^2I_{zz}}{dt^2}\sin^2 \theta,
\label{eq:quad}
\end{equation}
where $I_{zz}$ is the reduced-mass quadrupole moment, $G$ is the gravitational constant, $c$ is the speed of light, $D$ is the distance to the source (our fiducial value is $D=10$ kpc), and $\theta$ is the latitudinal angle between the supernova axis of rotation and the observer.  For extracting the GW signal from our 3D simulations we follow the method outlined in \citet{oohara:1997}.  For our analysis, we assume optimal source orientation---GWs emitted from the equator of the CCSN ($\theta = \pi/2$).  Later, as we outline our method to constrain progenitor $\xi_M$, we will discuss the impact of source orientation.

When analyzing the frequency structure of the GW signal, the \textit{peak} GW frequency is often a quantity of interest.  In this work, for our axisymmetric simulations, we use a similar form of the semianalytic formula proposed by \citet{muller:2013}

\begin{equation}
    f_\mathrm{peak} \sim \frac{1}{2\pi}\frac{GM}{R^2 c}\sqrt{2.1\frac{m_n}{\langle E_{\Bar{\nu}_e}\rangle}}\Bigg(1 - \frac{GM}{Rc^2}\Bigg)^2 ,
    \label{eq:fpeak}
\end{equation}
where $M$ is the mass of the PNS, $R$ is the PNS radius, $\langle E_{\Bar{\nu}_e}\rangle$ is the mean electron antineutrino energy, $G$ is Newton's gravitational constant, $c$ is the speed of light, and $m_n$ is the mass of a neutron.  (Note the factor of 2.1, instead of 1.1 in the original work).  While this may correspond to a physically higher adiabatic index approximating the pressure of the baryons near the PNS layer, we find this new factor reproduces the peak GW frequencies better for our axisymmetric simulations.  For the 3D models s40o[0-0.5]$^{3D}$ we use a similar form (Eqn. (5) of \citet{pan:2018}) that does not have the last quadratic term $(1 - GM/Rc^2)^2$ because the IDSA neutrino treatment does not account for gravitational redshift.

%

\subsection{Quality of Fit}

As we analyze over 30 multidimensional simulations over a wide range of parameter space, we perform functional fits to our data.  To quantify how well our models fit the data we use $\texttt{curve\_fit}$ in the $\texttt{scipy.optimize}$ library to return best-fit parameters as well as the standard deviation errors for each parameter.  Furthermore, we calculate the coefficient of determination as

\begin{equation}
    r_{\mathrm{det}}^2 = 1 - \frac{\Sigma_i (y_i - \hat{y}_i)^2}{\Sigma_i (y - \bar{y}_i)^2},
\label{eq:r2}
\end{equation}
where $(y_i -\hat{y}_i)^2 $ represents the squared residual between the simulation output data and model fit \citep{hughes:1971}.  The quantity $(y - \bar{y}_i)^2$ represents the variance of the simulation output data.  Here, $r_{\mathrm{det}}^2 = 0$ indicates no correlation with the regression line and $r_{\mathrm{det}}^2 = 1$ indicates a perfect correlation.  We use $r_{\mathrm{det}}^2$ to quantify the regression of our fit coefficients seen in the polynomial fits of Figures \ref{fig:hb_vs_J}, \ref{fig:fdot_vs_xi}, \ref{fig:fdot_vs_J}, and \ref{fig:3D_param}.

\section{Results}
\label{sec:results}

Our analysis synthesizes multiple components of the GW signal to constrain supernova progenitor compactness at collapse.  This project builds upon previous works that show the ramp-up slope of a nonrotating CCSN correlates with progenitor compactness \citep{warren:2020} and that the core-bounce signal encodes core angular momentum information \citep{abdik:2014}.  By ramp-up slope, we mean the slope of the peak GW frequency emitted in frequency versus time space.  The novel approach we take quantifies the \textit{rotational flattening}---or the decreasing slope of the ramp-up with increasing rotation rate---mentioned in \citet{pajkos:2019}.  Specifically, we find that the flattening correlates with angular momentum of the inner $1.75\, M_\odot$ of the supernova.  We then combine these two distinct features of the GW signal---loud bounce signal and dominant ramp-up slope---to constrain the core compactness of the progenitor star at collapse.  We first discuss the general evolution of each simulation, outline in detail each step of the analysis, and consider observability as well.  

\subsection{Evolution of Shock Radius}

\begin{figure}
    \centering
    \includegraphics[width=0.5\textwidth]{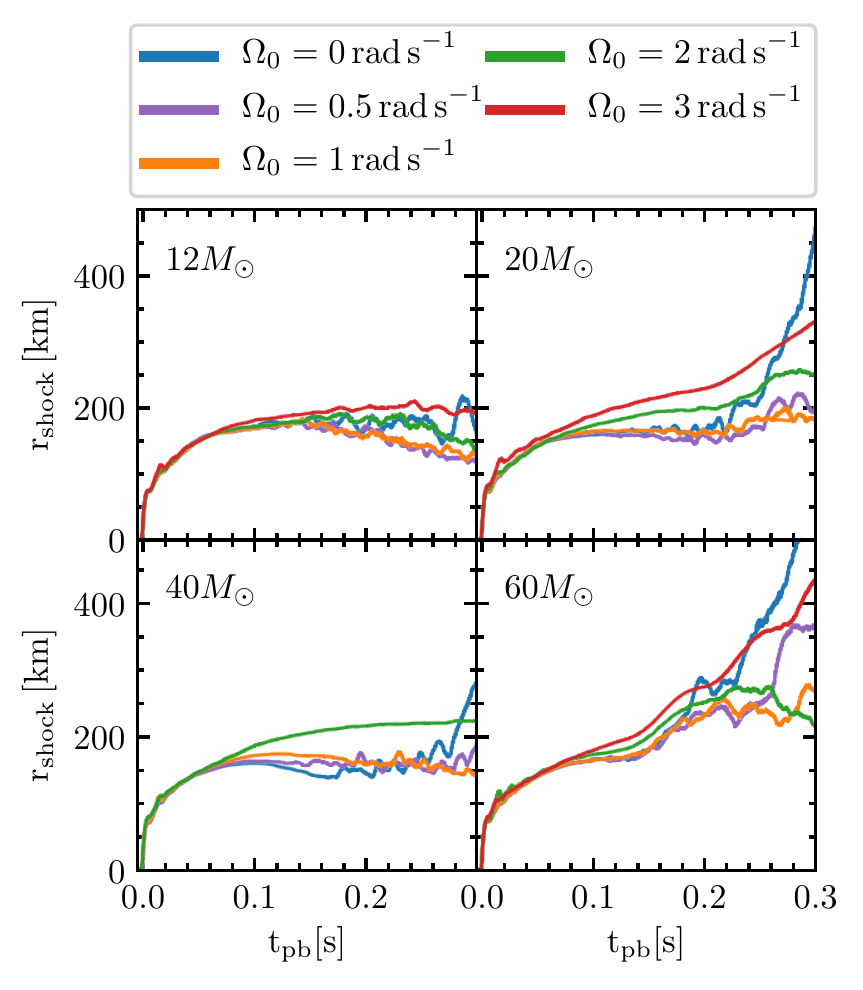}
    \caption{Evolution of the average shock radii for 2D all simulations using the SFHo EOS: models s12o[0-3], s20o[0-3], s40o[0-2], and s60o[0-3].  Similar to \citet{pajkos:2019}, we notice non-monotonic behavior of the shock expansion when considering progenitor mass and initial rotation rate.}
    \label{fig:shock}
\end{figure}


One of the main parameters we vary in this study is the central rotation rate, $\Omega_0$, so it is important to discuss the potential impact it can have on shock radius evolution.  On one hand, rotation can stabilize turbulence in the post-shock region, weakening one the main sources that drive a successful supernova explosion \citep{couch:2015a, janka:2016}.  Likewise, a centrifugally supported PNS that forms at larger radii would result in a softer neutrino spectrum.  These less energetic neutrinos could result in reduced heating behind the shock.  

On the other hand, rotation can also provide centrifugal support to a collapsing star.  This support would allow core bounce to occur at lower central densities, forming a shock at larger radii.  This initially, less gravitationally bound shock could be more conducive to a successful explosion.  Once again, we are reminded of the strong nonlinearities present when simulating CCSNe and the persistence of Mazurek's law: while changing one piece of physics may increase the likelihood of explosion, another factor is likely to change equally as much to counter that effect \citep{lattimer:2000}.

Here, we report the tendency of our 2D suite of models to explode.  Figure \ref{fig:shock} displays how the average shock radii of the supernovae evolve with time, with each panel dedicated to a specific progenitor mass.  The specific models plotted are s12o[0-3], s20o[0-3], s40o[0-2], and s60o[0-3].  Here, we define a simulation to \textit{successfully} explode if its averaged shock radius passes and remains 400 km away from the supernova center.
All models that do so also obtain a substantially positive diagnostic explosion energy \citep{bruenn:2016}. 

The $12 \,M_\odot$ progenitor at all rotational velocities does not undergo any successful explosions within the 300 ms simulation window.  It seems neither the supportive nor inhibitive nature of rotation drastically modifies the shock evolution.  
This is in agreement with previous results using FLASH-M1 for simulations of a nonrotating 12 $M_\odot$ progenitor that also failed to find explosions \citep{oconnor:2018} and the 2D results using the FORNAX code \citep{vartanyan:2018}. 
\citet{burrows:2019}, however, find a successful explosion for a nonrotating $12\, M_\odot$ progenitor in 3D, as do \citet{summa:2016} in 2D using the PROMETHEUS-VERTEX code.

The $20\, M_\odot$ explodes for the nonrotating case with one of the most aggressively advancing shock radii of the simulation set.  For $\Omega_0 = 3$ rad s$^{-1}$, while it does not reach 400 km within the simulation time, it does display steady growth to larger radii.  The simulations with the remaining rotational velocities show no further shock expansion.  

The four simulations of the $40\, M_\odot$ progenitor also do not successfully explode.  Indeed, it seems the high amounts of angular momentum endowed to this highly compact progenitor have a negative effect on the advancing shock front.  While all rotating cases show no significant shock displacement, it is worth noting that by the end of the simulation, the nonrotating case shows significant increase in its \textit{rate} of shock expansion.  Perhaps, with longer simulation times, the nonrotating $40\, M_\odot$ could show signs of exploding.

The $60\, M_\odot$ shows the most diverse behavior of the simulation set.  While the $\Omega_0 = 1, 2$ rad s$^{-1}$ simulations remain roughly stagnant at 300 ms pb, the remaining simulations advance toward or beyond 400 km.  Interestingly---similar to the $20\, M_\odot$ case---the nonrotating and fastest rotating models have the least bound shock radii.  Clearly, for the density profile within the $60\, M_\odot$ progenitor, the nonlinear effects of rotation become apparent.

While analyzing the \textit{explodability} of different models in detail is a key component to CCSN research, it lies beyond the scope of this paper, which focuses on the GW signals emitted.  Seconds after the core bounce, when the asymptotic explosion energy approaches a final value, the matter distribution and net neutrino production can be asymmetric.  In certain cases, these asymmetries can produce a direct, non-oscillatory, GW signal (sometimes referred to as a memory component) \citep{vartanyan:2020}.  While some of our models have significantly expanding shock radii, our simulations do not evolve to late enough times to develop the asymmetries necessary to produce this direct GW offset.

\subsection{General Features of Rotating GW Signals}

\begin{figure}
    \centering
    \includegraphics[width=0.5\textwidth,trim={20 0 0 0}]{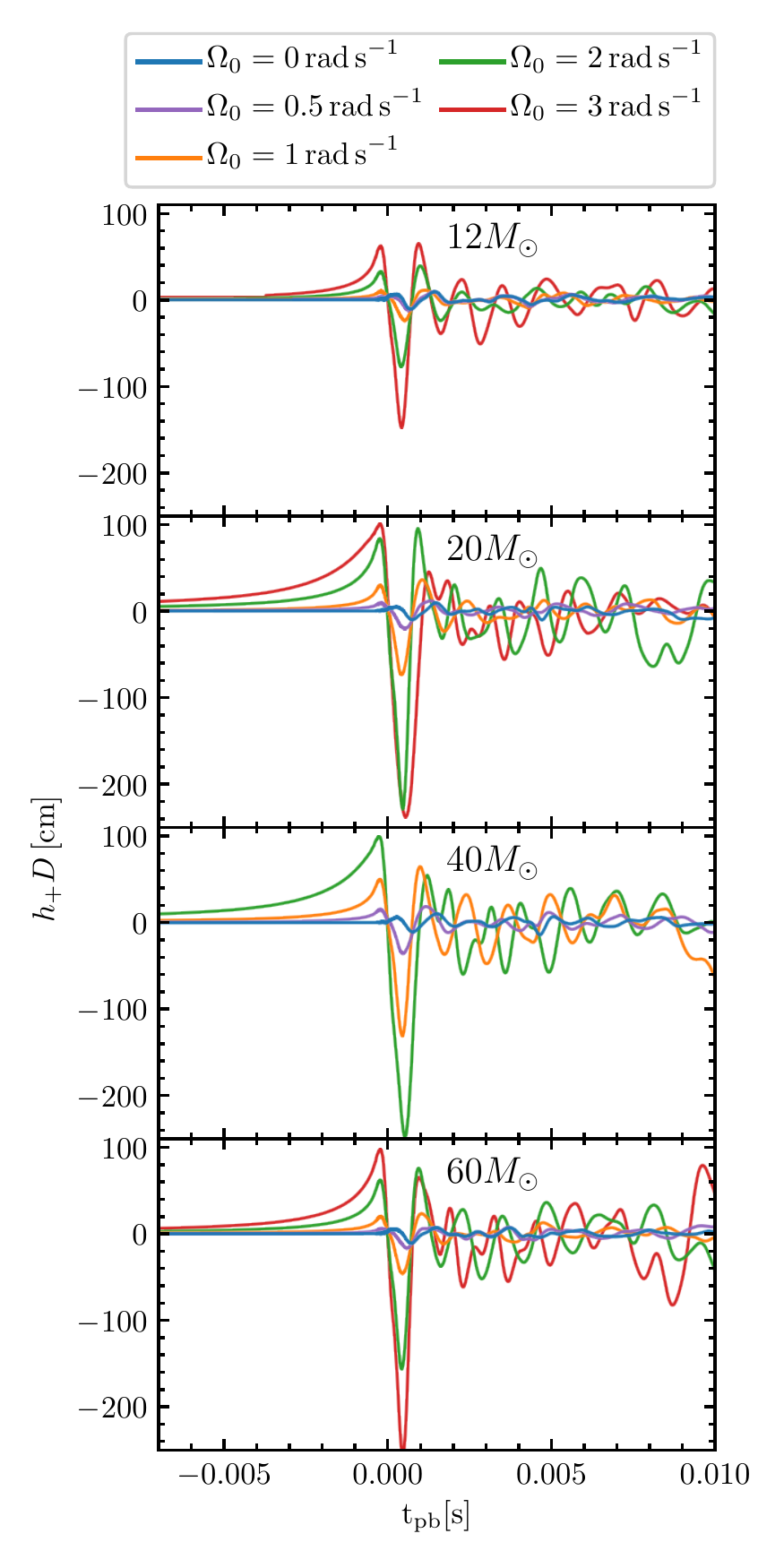}
    \caption{Bounce signals for all 2D CCSNe models with the SFHo EOS: models s12o[0-3], s20o[0-3], s40o[0-2], and s60o[0-3].  (Assumed distance of 10 kpc.)}
    \label{fig:TDWF_bounce}
\end{figure}

\begin{figure*}
    \centering
    \includegraphics{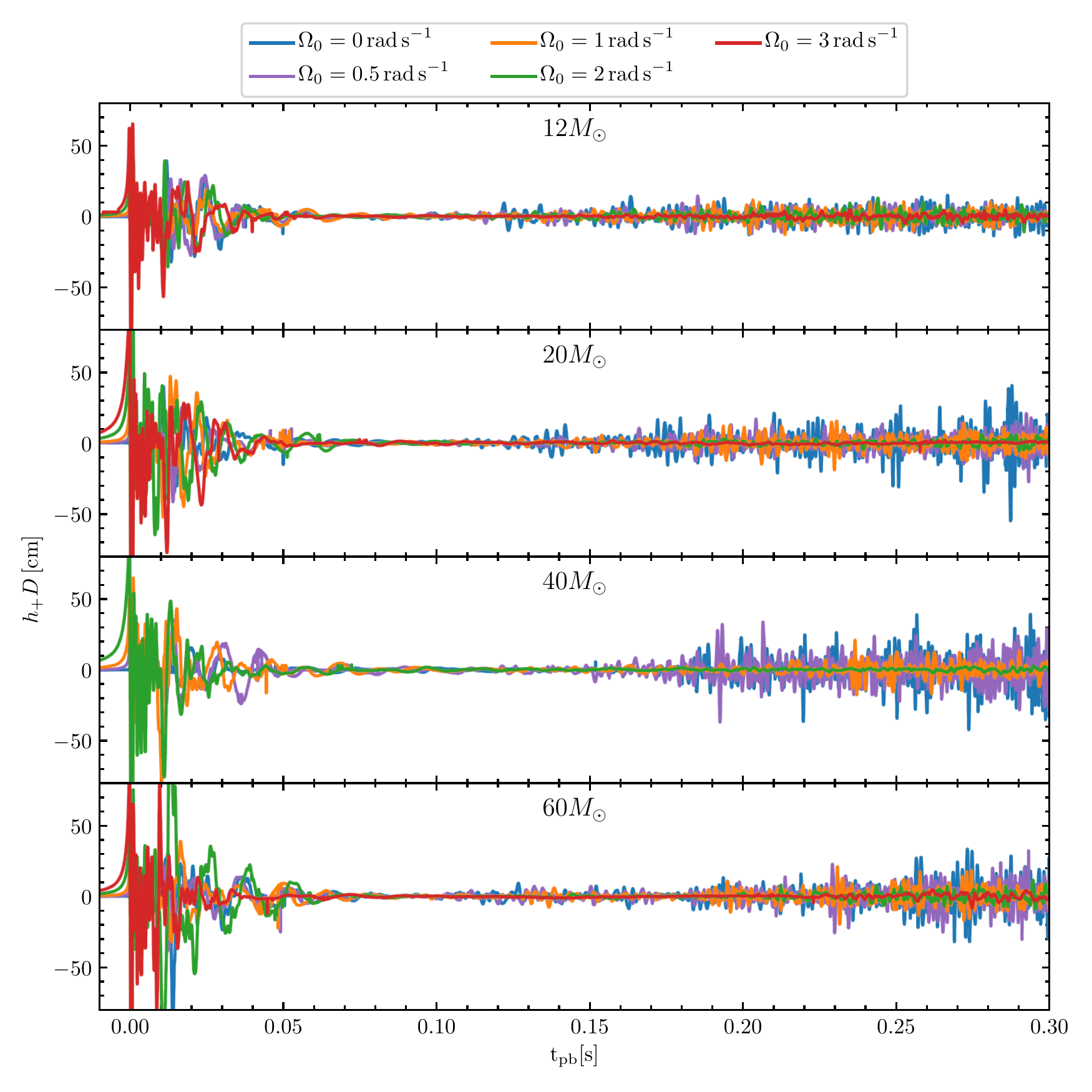}
    \caption{Time domain waveforms for all progenitors using the SFHo EOS: models s12o[0-3], s20o[0-3], s40o[0-2], and s60o[0-3]. (Assumed distance of 10 kpc.)  Two universal trends seen across our parameter space is that the bounce amplitude increase and the accretion phase GW amplitude decrease, with increasing rotational velocity.  In certain rotating cases, the presence of strong prompt convective activity also leads to high amplitude GWs 10s of ms after bounce.}
    \label{fig:TDWF_all}
\end{figure*}

Here, we review the general features of the GW signal seen in our rotating and nonrotating 2D simulations using the SFHo EOS.  Figure \ref{fig:TDWF_bounce} shows the bounce signals from all rotation rates, separated by progenitor mass, while Figure \ref{fig:TDWF_all} displays the waveforms over the entire simulation duration.  Specifically, the models included in these figures are s12o[0-3], s20o[0-3], s40o[0-2], and s60o[0-3].

The first main feature is the loud bounce signal.  As rotation rate increases, the increasing angular momentum of the once-stellar iron core forces it to deviate from spherical symmetry to become an oblate spheroid.  At bounce, the nuclear force suddenly halts the infalling matter causing the (once iron core, now) PNS to deform on the time scale of a fraction of a millisecond.  More specifically, to leading order, the mass quadrupole moment of the PNS drastically accelerates, resulting in the emission of GWs.  In extreme scenarios, for sufficient angular momentum within the inner core, the infalling matter can become centrifugally supported.  This slowed infall, in turn, deforms the PNS less at bounce, creating a GW bounce signal with a lower amplitude \citep{dimm:2008}.

In Figure \ref{fig:TDWF_bounce}, the GW strain the first few milliseconds after bounce corroborates previous findings that describe, except in these extreme cases, that increasing the angular momentum of the core will yield a larger bounce signal \citep{dimm:2008}.  As this bounce signal is largely governed by the microphysics within the core, the GW bounce signal is predictable and well templated \citep{scheidegger:2010b}, provided a set of assumptions about the microphysics.  
As the PNS rings down from the energetic bounce, matter motions due to prompt convection contribute to the GW signal 10s of ms after bounce as well.

The accretion phase of the supernova---the hundreds of milliseconds when the PNS is accreting matter---marks the next time during the supernova when significant GWs are produced.  During accretion, infalling stellar material and convection in the gain layer can excite oscillatory modes in the PNS.  As a result, material at nuclear densities moves on time scales the order of milliseconds and can result in sustained GW emission.  For all of our rotating progenitors, we observe this GW signal occurring between $\sim 100-300$ ms pb in Figure \ref{fig:TDWF_all}.  In contrast to the bounce signal, these oscillatory modes are largely stochastic in nature and nearly impossible to template when viewed in the time domain. 

Another interesting feature during the accretion phase, across all progenitors, is the decrease in amplitude of the GW signal with increasing rotation rate.  As pointed out in previous work, this \textit{rotational muting} is a direct consequence of the Solberg-Hoiland stability criterion, i.e., the stabilizing effect rotation can have in a convective fluid \citep{endal:1978,pajkos:2019}.  Simply put, in these 2D simulations, as rotation rate increases, convective activity slows in the post-shock region of the supernova.  This slowed convection will interact less with the PNS---causing less pronounced oscillations---resulting in lower amplitude GWs.  Of course, 3D rotational instabilities like the spiral mode of the SASI \citep{andresen:2019}, the low $T/|W|$ instability \citep{pan:2020,shibagaki:2021} or magnetorotational instability (MRI) \citep{akiyama:2003,cerda-duran:2007,mosta:2015} can create turbulent motion as well, re-exciting the motion of the PNS.  Nevertheless, it is promising that rotational muting is still observed in our simulations with an increased fidelity in microphysics---namely, inelastic electron scattering and velocity dependence in the neutrino transport---compared to our previous works.\newline \\


\subsection{Connecting Angular Momentum and the Bounce Signal}
\label{subsec:bounce}

We now move on to laying out a novel analysis method that will help constrain the rotational information of the supernova at the time of core bounce and mass distribution within the progenitor star at collapse.  The first vital piece of information needed is the amplitude of the core-bounce GW signal from observations of a nearby CCSN.

It is well established that the source of the GW bounce signal is the dense supernova core, the region where the previous stellar iron core is forming into a PNS.  For rotating CCSNe, the amplitude of this bounce signal is well studied and has been shown to correlate with the ratio of rotational kinetic energy to gravitational potential energy--commonly displayed as $\beta \equiv T/|W|$ \citep{dimm:2008}.  While $\beta$ can be used to measure the degree of rotation in a supernova, we choose the angular momentum instead, in aims to establish it as another key quantity, when quantifying rotation within a CCSN.  Nevertheless, these quantities are related, as they are both metrics that encode information about the supernova rotation profile and mass distribution.  For scale, the corresponding upper value in our simulation suite is  $J^\mathrm{bounce}_{1.75M_\odot} \sim 2.8 \times 10^{49}$ erg s.  The upper limit for the core of the CCSN is $\beta^\mathrm{bounce}_\mathrm{core} \sim 0.10$.  Our definition of supernova core is outlined at the end of this subsection.

\begin{figure}
    \centering
    \includegraphics[width=0.5\textwidth]{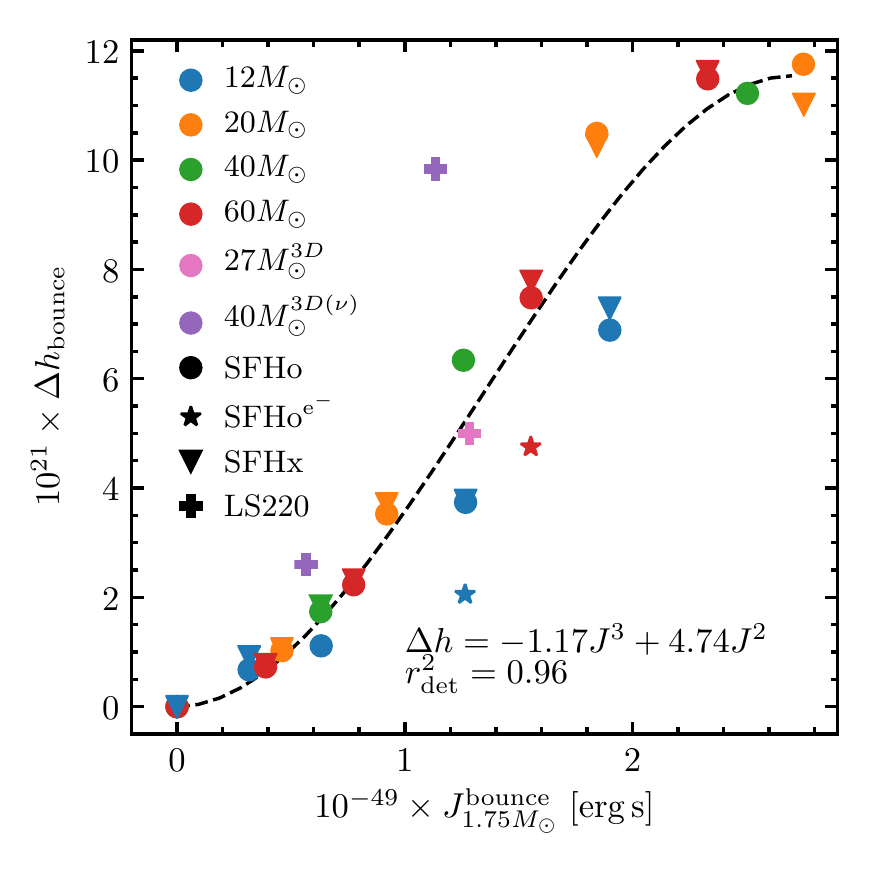}
    \caption{Bounce signal amplitude versus angular momentum of inner $1.75 \,M_\odot$ for all models in our simulation suite.  Overlaid is a third order polynomial fit to the 33 models using the M1 neutrino treatment and the SNA approximation \citep{bruenn:1985} when calculating neutrino opacity tables.   Colors represent different progenitor masses, whereas the different shapes correspond to the specified EOS listed by each black legend marker. This correlation is nearly EOS independent.  Every shape uses the SNA approximation \citep{bruenn:1985} when calculating neutrino opacity tables, with the exception of the stars ($\bigstar$).  Stars (labeled SFHo$^{e^-}$) use LMP+N50 $e^-$ capture rates which affect the deleptonization of the core during collapse, resulting in a lower amplitude bounce signal \citep{langanke:2001,titus:2018}.  Specifically, the two models that use the LMP+N50 $e^-$ capture rates correspond to s12o2$^\nu$ and s60o2$^\nu$.  Likewise, the difference in neutrino treatment used in the $40 M_\odot^{3D(\nu)}$ models (s40o0.5$^{3D}$ and s40o1$^{3D}$) causes deviation from the fit.}
    \label{fig:hb_vs_J}
\end{figure}

In Figure \ref{fig:hb_vs_J}, we relate the amplitude of the bounce signal ($\Delta h_{\mathrm{bounce}}$) to the angular momentum of the inner $1.75 \,M_\odot$ at bounce, by performing a third order polynomial fit to the 33 simulations using the M1 neutrino treatment and single nucleus approximation (SNA) $e^-$ capture rates \citep{bruenn:1985}, or all models \textit{except} $s12o2^\nu$, s$60o2^\nu$, and s40o[0-1]$^{3D}$.  In this work, we define the $\Delta h_{\mathrm{bounce}}$ as the difference between the maximum and minimum of the strain within a 3 ms window of the bounce time---the time when the central entropy reaches 3 $k_\mathrm{B}$ baryon$^{-1}$ and central density exceeds $2\times 10^{14}$ g cm$^{-3}$.  Our mass cut of $1.75\, M_\odot$---instead of typical values $\sim 0.6 \,M_\odot$ for inner cores---is chosen in anticipation of our analysis that will contain information about the accreted matter, beyond the canonical PNS.  A more detailed justification of this mass cut is discussed in Section \ref{subsec:rot_flat}.

We now move to outlining the two main features of Figure \ref{fig:hb_vs_J}:  the quadratic $J$ regime and the extreme asymptotic $J$ regime.  To explain the quadratic $J$ regime, when $J_{1.75M_\odot} \lesssim 2 \times 10^{49}\, \mathrm{erg\, s}$, we appeal to order of magnitude estimates of the bounce signal.  As established in \citep{richers:2017}, the bounce signal can be approximated as $\Delta h\sim GMR^2\Omega^2/c^4D$ for a given PNS mass $M$, radius $R$, and rotation rate $\Omega$.  The angular momentum at the center of the supernova just after core bounce can be approximated as $J\sim MR^2\Omega$.  This yields an expression for how $\Delta h$ depends on $J$,

\begin{figure}
    \centering
    \includegraphics[width=0.5\textwidth]{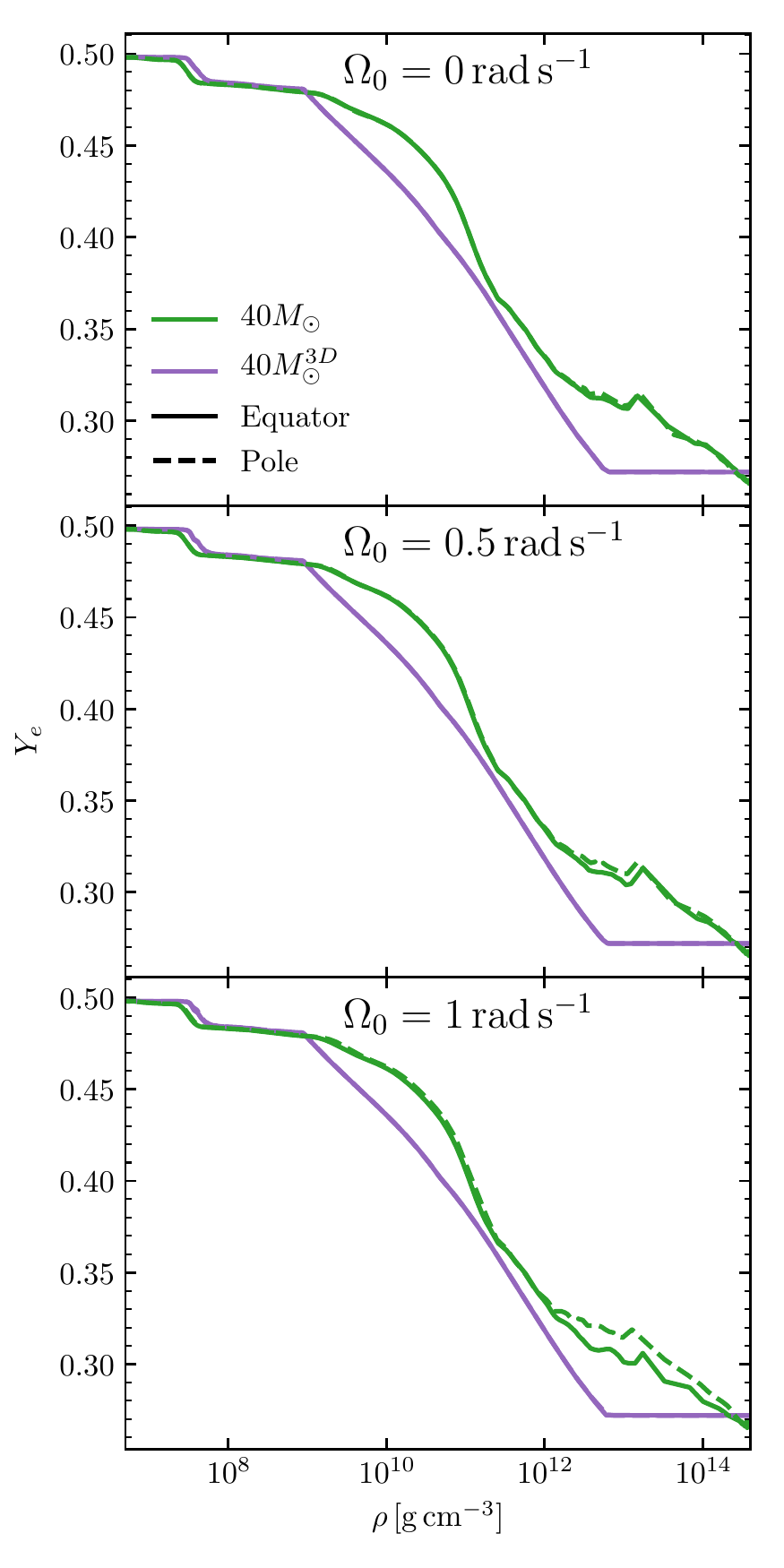}
    \caption{ Electron fraction vs. density ($Y_e(\rho)$) profiles for model s40o[0-1] and s40o[0-1]$^{3D}$, which use M1 and parameterized deleptonization \citep{lieb:2005} on collapse, respectively.  Due to the different treatments of deleptonization on collapse, both cases have extremely similar central $Y_e$ ($Y_e^c$) values of $\sim 0.27$ yet produce differences in $\Delta h_{\mathrm{bounce}}$ of up to $4\times 10^{-21}$.  Because of this evidence, we caution against using $Y_e^c$ as a predictor for $\Delta h_{\mathrm{bounce}}$.  Rather, some metric involving the $Y_e$ over the entire radius of the PNS would be preferred.  Examining the s40o[0-1] models (green), we notice an angular dependence of the $Y_e(\rho)$ profiles (solid vs. dashed lines), particularly for $\Omega_0 = 1$ rad s$^{-1}$.  This difference stems from the velocity dependence of the neutrino transport in the M1 scheme, which is affected by changes in the radial velocity field configuration, induced by rapid rotation.  Furthermore, it provides evidence against the assumption that deleptonization on collapse is a spherically symmetric process, particularly in rapidly rotating progenitors.}
    \label{fig:ye_vs_rho}
\end{figure}

\begin{equation}
    \Delta h \sim \frac{G}{c^4DMR^2}J^2.
    \label{eq:h_vs_J}
\end{equation}

Thus, for typical $M$ and $R$ values of the supernova center ($\sim 0.55 M_\odot$ and $\sim 15$ km in our simulations), one notes the quadratic behavior seen in Figure \ref{fig:hb_vs_J} for $J_{1.75M_\odot} \lesssim 2\times 10^{49}$ erg s.  Of course this scaling relation is constructed for the main source of GWs, the PNS, or inner $\sim$ 0.6 $M_\odot$ of matter.  Nevertheless, we find this quadratic behavior maps quite well to larger mass cuts, as seen in Figure \ref{fig:hb_vs_J}.

For $J_{1.75M_\odot} \gtrsim 2 \times 10^{49}\, \mathrm{erg\, s}$, the supernova center enters the extreme asymptotic $J$ regime.  The supernova center now contains sufficient $J$ to centrifugally support the collapse and causes the bounce signal to occur at lower densities.  This effect in turn widens the bounce signal and prevents further growth of $\Delta h_{\mathrm{bounce}}$ \citep{fryer:2004, dimm:2008}.

Another interesting feature of Figure \ref{fig:hb_vs_J} is the independence of $\Delta h_{\mathrm{bounce}}$ with EOS, which corroborates previous works  \citep{dimm:2008,richers:2017}.  Denoted circles are simulations that use the SFHo EOS: s12o[0-3], s20o[0.5-3], s40o[0.5-2], and s60o[0.5-3].  Inverted triangles use the SFHx EOS: s12o[0.5,2,3]x, s20o[0.5-3]x, s40o0.5x, and s60o[0.5-3]x.  Over this parameter space, both EOSs produce similar bounce amplitudes.  While we have not exhausted the possible list of available EOSs, the fact that two distinctly different EOSs produce similar bounce signals is promising to apply an analysis of this type to future GW observations and highlights the supernova mass distribution as a contributing factor to determining the bounce amplitude.

For the 3D simulations, which all use the LS220 EOS, we notice good agreement with the 3D 27 $M_\odot$ case, s27o2$^{3D}$.  However, for the rotating 40 $M_\odot$ simulations (s40o0.5$^{3D}$ and s40o1$^{3D}$) we notice systematically higher bounce signals that deviate from the polynomial fit in Figure \ref{fig:hb_vs_J}.  This difference can be attributed to the neutrino treatments in each simulation suite.  For all 2D and the 27 $M_\odot$ 3D cases, M1 neutrino transport is used beginning at collapse.  This treatment gives accurate deleptonization behavior within the supernova center by the time of core bounce.  Models s40o0.5$^{3D}$ and s40o1$^{3D}$, by contrast, use parameterized deleptonization \citep{lieb:2005} up until the point of bounce.  At densities above $10^9$ g cm$^{-3}$ these two schemes yield different electron fraction versus density, or $Y_e(\rho)$, profiles at the time of bounce.  As explained in \citet{richers:2017}, this effect in turn modifies the amplitude of the bounce signal. 

Another piece of physics that impacts the bounce signal significantly are the electron capture rates.  We perform two additional simulations with identical initial conditions to the 12 and 60 $M_\odot$ progenitors with $\Omega_0 = 2$ rad s$^{-1}$ (s12o2$^\nu$ and s60o2$^\nu$), except for different neutrino opacity tables.  The control case used for our 2D simulation suite creates the neutrino interaction library for the SFHo EOS using $\texttt{NuLib}$ \citep{oconnor:2015} and uses the SNA approximation \citep{bruenn:1985}.  To test the effect of modified electron capture rates, we use the weak rate library of Laganke and Martinez-Pinedo  \citep{langanke:2001} supplemented by the calculations of \citet{titus:2018}.  Between the two cases, we notice bounce signal amplitudes that differ by $\sim 40$\%.  Similar to before, this difference in bounce signal is expected.  As core collapse commences, electron captures onto nuclei play a significant role in removing pressure support from the iron core.  Eventually, at the time of bounce, this will modify the $Y_e(\rho)$ profile, thereby affecting the core mass and resulting GW bounce signal \citep{richers:2017}.  Because of this clear dependence of the GW bounce signal on electron capture rates, this work serves as valuable scientific motivation to reduce experimental error on these rates in high density nuclear matter, as they have an impact on our ability to constrain supernova progenitor information from GWs alone.

In many of the primary works that examine GWs from rotating CCSNe, correlations between inner core mass at bounce ($M^B_\mathrm{core}$) and central electron fraction ($Y_e^\mathrm{c}$) are treated as diagnostics for the expected $\beta^\mathrm{bounce}_\mathrm{core}$ (and consequently $\Delta h_\mathrm{bounce}$). \citep{dimm:2008,scheidegger:2010,abdik:2014,richers:2017}.  One dominant common factor between these works is the inclusion of parameterized deleptonization on collapse \citep{lieb:2005}.  In our work, over the range in values for $0 < \beta^\mathrm{bounce}_\mathrm{core} \lesssim 0.10$, we do not see significant changes in values of $M^B_\mathrm{core}$--- changes less than a 0.1 $M_\odot$.  Likewise, the common correlations of increasing $Y^\mathrm{c}_e$ with increasing $\Delta h_\mathrm{bounce}$ are not completely upheld with our data.  As an example, refer to Figure \ref{fig:ye_vs_rho}.  When comparing models s40o1 and s40o1$^{3D}$, the principle difference during collapse is the neutrino treatment: s40o1 uses velocity-dependent M1 neutrino transport, whereas s40o1$^{3D}$ uses parameterized deleptonization.  While both have extremely similar $Y^\mathrm{c}_e$ values, the $\Delta h_\mathrm{bounce}$ of s40o1 is only $\sim 60\%$ that of s40o1$^{3D}$.  To explain this difference, we examine the $Y_e(\rho)$ profile at bounce.   We observe that at densities greater than $10^{12}$ g cm$^{-3}$ the difference in profiles can vary by as much as $\Delta Y_e \sim 0.02$ (around 6$\%$), while still maintaining similar $Y^\mathrm{c}_e$.  While we acknowledge the value in the 1D Boltzmann neutrino transport calculations involved in \citet{lieb:2005}, we present evidence using our multidimensional velocity-dependent neutrino-radiation hydrodynamic simulations that the inclusion of rotation and multidimensional effects can cause this approximation to break down, and recommend caution for use in future rapidly rotating GW studies.

To define $M^B_\mathrm{core}$ in our work, we look at the magnitude of the velocity along the pole and equator.  Where each radial profile has the steepest velocity gradient, we select corresponding points, as this marks the point where sonic contact breaks down.  Lastly, we construct an ellipse using these corresponding radial points as the semimajor and semiminor axes.  All mass within this ellipse we define as \textit{in sonic contact} with the supernova center.  

As a caveat, it is important to note that these bounce signals were calculated with an assumed distance of 10 kpc.  In the event of an actual rotating CCSN event, we would rely on other means to constrain the distance.  Of course, EM observations remain the gold standard.  For a purely multimessenger approach, distance estimates using supernova neutrino measurements still lie on the horizon.\newline 

\subsection{$\dot{f}$ versus Compactness}
\begin{figure}
    \centering
    \includegraphics[width=0.5\textwidth]{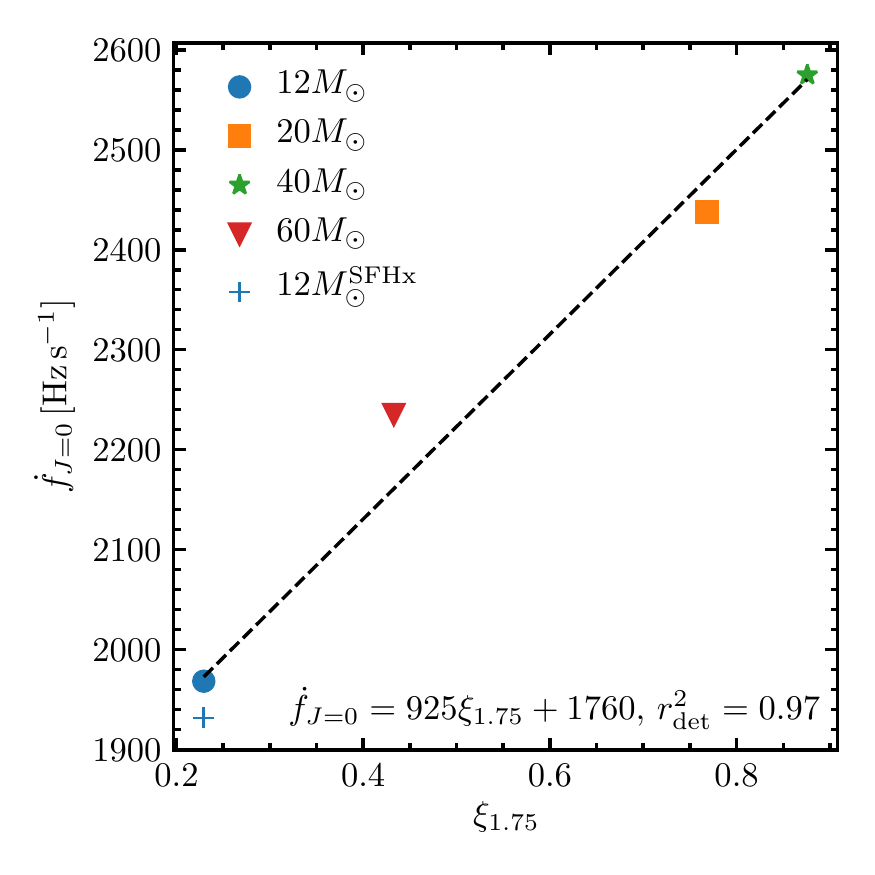}
    \caption{Slope of the PNS ramp-up in frequency vs. time space for nonrotating progenitors, as a function of $\xi_{1.75}$ at collapse.  These models correspond to s12o0, s20o0, s40o0, s60o0, and s12o0x.  More compact progenitor cores will yield higher mass accretion rates, leading to faster contraction of the PNS radius--in turn, steeper ramp-up slopes. }
    \label{fig:fdot_vs_xi}
\end{figure}

The next piece of information needed for this analysis requires us to connect a GW observable to the density profile of the supernova.  Through a variety of 1D simulations, \citet{warren:2020} recently explored the GW signal a few seconds after bounce, displaying correlations between progenitor compactness at collapse and the slope of the GW frequency ramp-up ($\dot{f}$), in frequency versus time space.  To calculate $\dot{f}$, we perform a linear regression to Equation (\ref{eq:fpeak}) between 50 and 300 ms pb.  We choose to begin tracking $f_\mathrm{peak}$ at 50 ms because this approximately marks the end of the post-bounce ring-down GW signal seen in Figure \ref{fig:TDWF_all}.  Because we are using multidimensional simulations and use a different mass cut for $\xi_M$ at collapse, we do not use the parametric fit provided in \citet{warren:2020}.  Instead, we opt to create a unique linear fit for our 2D simulations. Nevertheless, we corroborate a linear relationship between $\dot{f}$ and $\xi_M$ in 2D simulations.  Figure \ref{fig:fdot_vs_xi} displays $\dot{f}$ versus $\xi_{1.75}$ at collapse, for the nonrotating simulations in our suite: specifically, models s12o0, s20o0, s40o0, s60o0, and s12o0x.  When comparing the EOS dependence of ramp-up slopes for nonrotating progenitors, we note only a difference of $\sim 2\%$ between SFHo and SFHx.

Physically, this $\dot{f}$ and $\xi_M$ relationship should be expected.  At the onset of collapse, a higher compactness value corresponds to more mass closer to the stellar core.  As the inner part of the resulting supernova will be more gravitationally bound, the mass accretion rate onto the PNS should be higher, compared to a lower $\xi_M$ progenitor.  With a higher mass accretion rate, the PNS should contract on shorter time scales.  Thus, because the GW ramp-up is related to the PNS dynamical frequency \citep{camp:2004}, faster PNS contraction will lead to a higher $\dot{f}$ value.  

As a note, we acknowledge that empirically measuring $\dot{f}$ from the GW signal simulation outputs would be ideal.  However, the frequency components for the $\Omega_0 = 2,3$ rad s$^{-1}$ simulations have extremely faint GW signals during the accretion phase, which would leave an unreliable fit.  Physically, this weaker signal is a product of the \textit{rotational muting} seen in rotating CCSN simulations \citep{pajkos:2019}.  Because these simulations are axisymmetric, 3D effects such as the spiral mode of the SASI or low $T/|W|$ instabilities do not arise to re-excite the PNS oscillations \citep{andresen:2019}.  In the 3D context, as in nature, these instabilities can arise in rotating cases, creating a detectable GW signal in conjunction with a flatter $\dot{f}$.  The use of Equation (\ref{eq:fpeak}) has been shown in previous  works \citep[eg.,][]{muller:2013,pan:2018} to reproduce the peak GW frequency quite well for nonrotating CCSNe.  To test its effectiveness for rotating models, we show how it tracks the GW frequency output from our 2D simulations, in the next subsection.

The only model that does not use a semianalytic model, similar to Equation (\ref{eq:fpeak}), to calculate $\dot{f}$ is model s40o1$^{3D}$.  As noted in \citet{pan:2020}, 40o1$^{3D}$ displays the low T/W instability during the accretion phase.  Accompanying the instability is a bar-mode-like configuration of the rotating PNS \citep{ott:2005}.  Compared to the slight oblateness seen in our 2D rotating models, this bar mode deviates from spherical symmetry enough to create a lower fidelity prediction of the peak GW frequency.  Instead, we opt to empirically extract the peak GW frequencies to construct $\dot{f}$ for s40o1$^{3D}$.  While other GW signals have been associated with non-axisymmetric instabilities---the emission of quasiperiodic GWs $\sim 100$ Hz from the spiral SASI \citep{kuroda:2014,andresen:2017} or transient approximate kilohertz signal from the low $T/|W|$ instability \citep{ott:2005}---we note these signals do not interfere with the empirical calculation of $\dot{f}$ from the dominant oscillation mode produced by s40o1$^{3D}$.  While we acknowledge $\dot{f}$ values between models s40o1 and s40o1$^{3D}$ differ by $\sim 10\%$, we cannot conclude this difference in \textit{non-normalized} ramp-up slope is due to the presence of the low T/W instability because of the difference in respective neutrino treatments (M1 and IDSA) and progenitor structure (\citep{woosley:2007} and \citet{sukhbold:2016}).  For the reader interested in the exact details of the spectrogram for s40o1$^{3D}$, we direct them to Figure 9 of \citet{pan:2020}.


\subsection{Quantifying Rotational Flattening}
\label{subsec:rot_flat}

\begin{figure}
    \centering
    \includegraphics[width=0.5\textwidth,trim={0 0 0 0}]{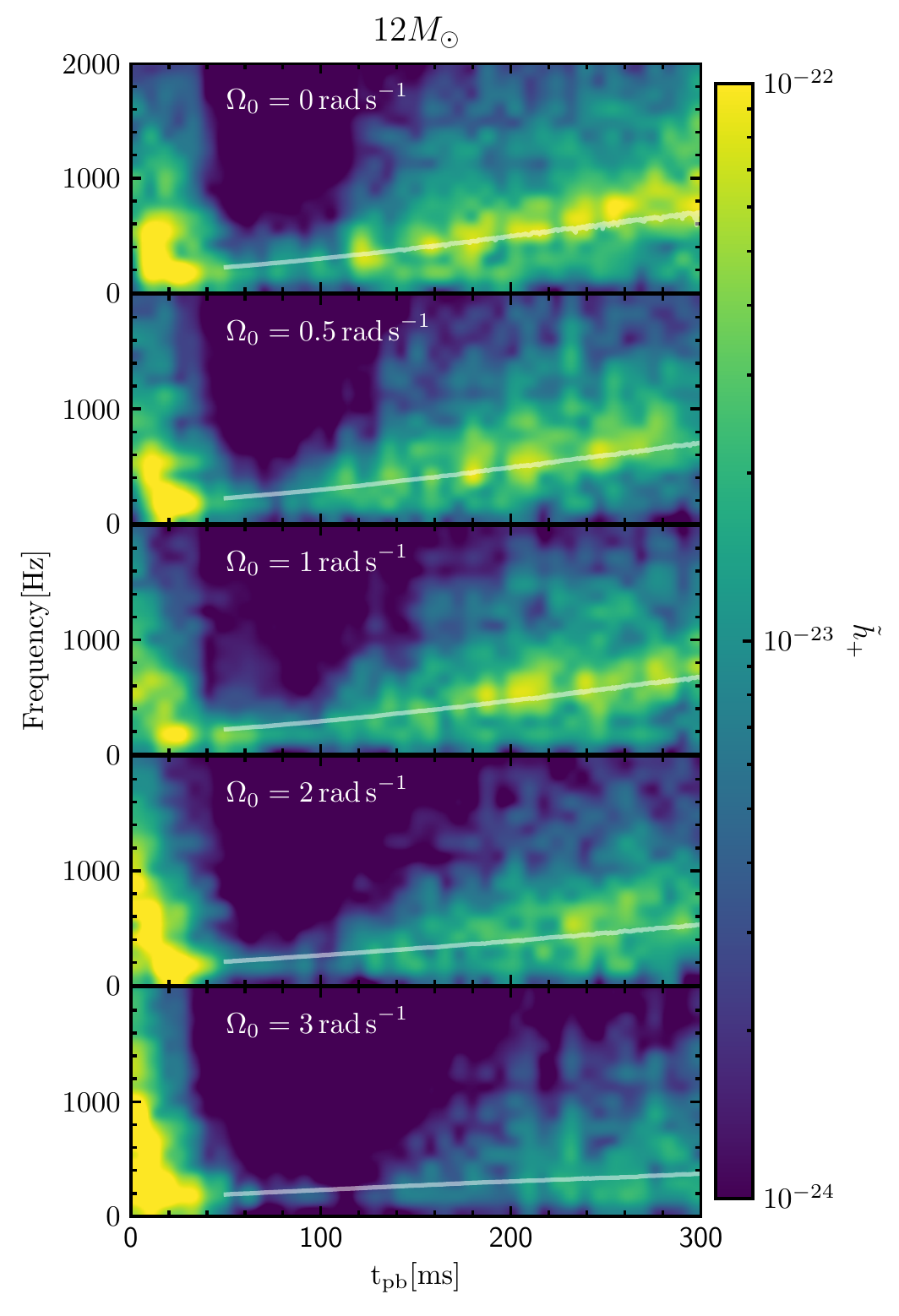}
    \caption{Spectrograms for the $12\,M_\odot$ progenitor at different rotational velocities: models s12o[0-3].  Overplotted in gray is the peak GW frequency of the PNS (Equation (\ref{eq:fpeak})), displaying the rotational flattening of the PNS, as the rotation rate increases.  The colors correspond to values of $\Tilde{h}_+$, the Fourier transform of the GW strain $h_+$.}
    \label{fig:spectrogram}
\end{figure}

\begin{figure}[t]
    \centering
    \includegraphics{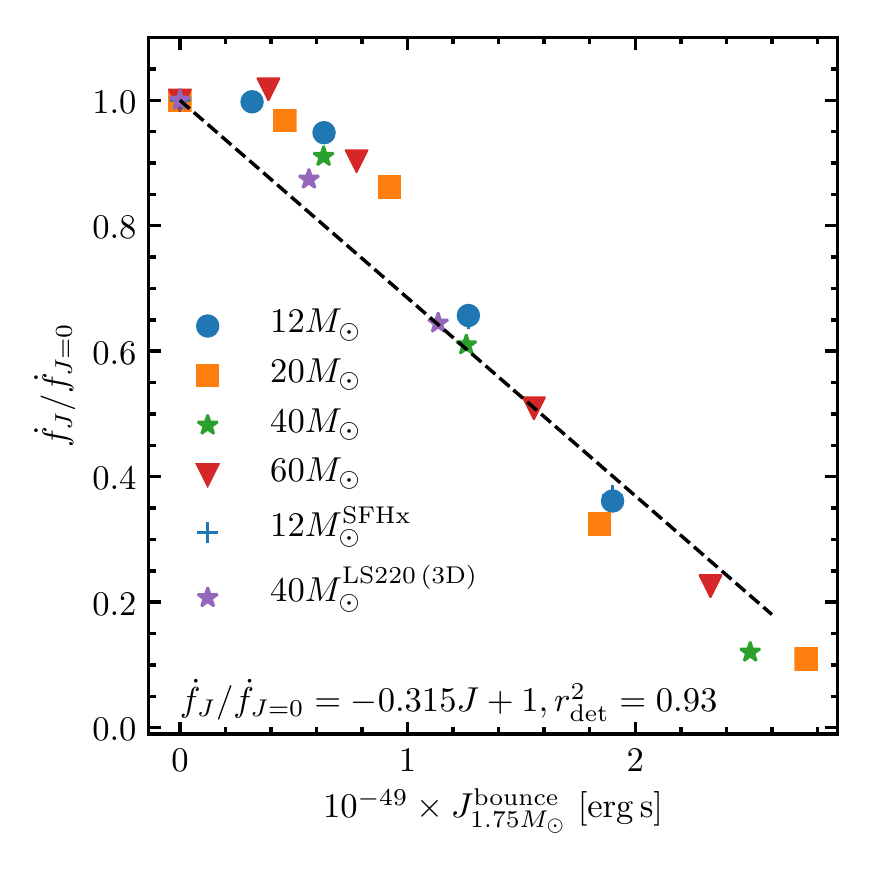}
    \caption{Linear correlation between how much the ramp-up slope is flattened vs. $J_{1.75M_\odot}$. We perform three additional simulations for the 12 $M_\odot$ progenitor using the SFHx EOS and see almost no EOS impact on the flattening effect.  This fact indicates the pre-accretion angular momentum distribution is more important when quantifying flattening.  In total, the models included in this figure are s12o[0-3], s20o[0-3], s40o[0-2], s60o[0-3], s12o[0,2,3]x, and s40o[0-1]$^{3D}$.}
    
    \label{fig:fdot_vs_J}
\end{figure}

\begin{figure*}
    \centering
    \includegraphics[width=\textwidth]{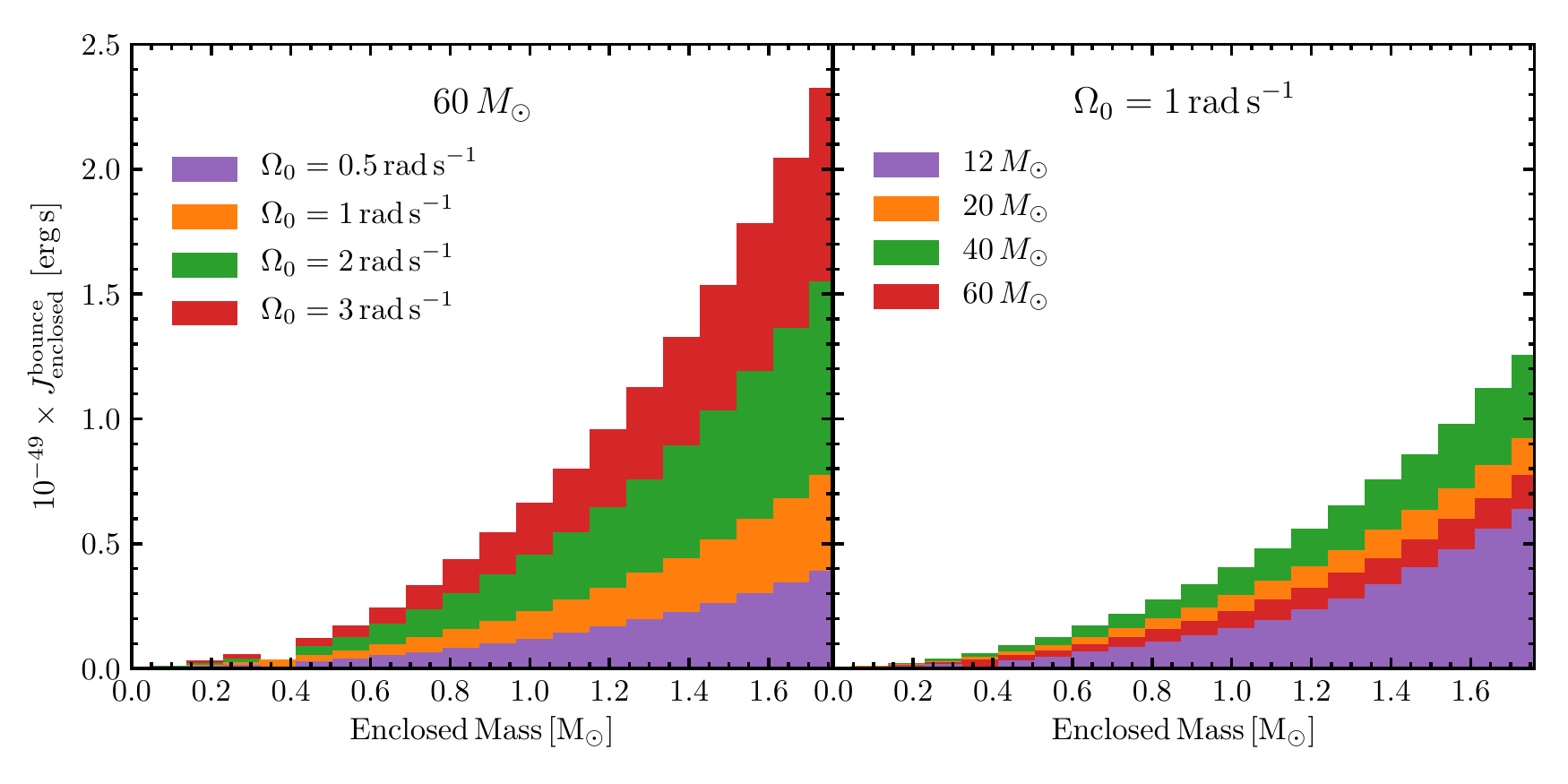}
    \caption{(Left) Enclosed angular momentum ($J_\mathrm{enclosed}$) binned by mass coordinate (20 total bins) for the $60 \,M_\odot$ progenitor at different rotation rates: models s60o0.5, s60o1, s60o2, and s60o3.  As rotation rate increases, the $J_\mathrm{enclosed}$ profile becomes steeper.  (Right) $J_\mathrm{enclosed}$ binned by mass coordinate (20 total bins) for all progenitor masses for $\Omega_0 = 1$ rad s$^{-1}$: models s12o1, s20o1, s40o1, and s60o1.  In this case, the $J_\mathrm{enclosed}$ profile becomes steeper with increasing compactness.  This is expected as the differential rotation parameter A used in Equation (\ref{eq:omega}) depends linearly on compactness.  Thus, the more compact models begin with larger angular momentum values at collapse.  (Both) In both cases, a steeper $J_\mathrm{enclosed}$ profile provides more angular momentum for the PNS to accrete after bounce---in turn, this provides more centrifugal support and a flatter $\dot{f}_J$.}
    \label{fig:Jprofile}
\end{figure*}

While the correlation between $\dot{f}$ and $\xi_M$ is indeed valuable, this relationship has only been shown for nonrotating cases thus far.  As all stars rotate to some degree, we now generalize this relationship beyond simple nonrotating cases, granting the final piece needed for our new analysis method: extracting rotational information from the accretion phase signal.

During the accretion phase of a CCSN, the PNS is accreting mass while cooling via neutrino emission.  These two factors cause the PNS to contract as the supernova evolves.  Observationally, this cooling manifests itself in the ramp-up slope of the GW signal.  Intuitively, a PNS with a smaller radius (or higher dynamical frequency) will oscillate at higher frequencies.  Thus, as the PNS radius gradually decreases, its frequency of emission should gradually increase.    However, if the PNS is rotating, it will receive centrifugal support during that cooling process.  Not only is the PNS accreting matter, but angular momentum from the overlying stellar material.  This accretion will \textit{spin up} the PNS, allowing it to end with a larger radius, compared to the nonrotating case.  With a larger radius (or smaller dynamical frequency), one expects the GW frequency to be lower.  If one were to observe the GW evolution in the time-frequency domain, the rotating PNS would appear to have an $\dot{f}$ that is \textit{flatter} (closer to 0) than the nonrotating case.  Now we quantify this \textit{rotational flattening} by examining the ramp-up slopes of different progenitors.  

In Figure \ref{fig:spectrogram}, we display spectrograms for the five $12 \,M_\odot$ simulations for various rotation rates: models s12o[0-3].  Brighter hues correspond to greater contributions to the GW signal at a given frequency.  Overlaid in gray is the $f_\mathrm{peak}$ produced by Equation (\ref{eq:fpeak}).  Indeed, with increasing rotation rate, $f_\mathrm{peak}$ evolves with a flatter slope.  Figure \ref{fig:spectrogram} shows how tightly $f_{\mathrm{peak}}$ tracks the frequency evolution of the emitted GWs.

After calculating $\dot{f}$ for each simulation, we normalize by the respective nonrotating ramp-up slope.  Interestingly, we find a tight correlation between how much the slopes are flattened over time and the angular momentum of the inner $1.75 \,M_\odot$ at bounce.  Figure \ref{fig:fdot_vs_J} displays this linear fit to the 19 2D SFHo runs.  For comparison, we overlay the three 2D SFHx runs and three 3D LS220 runs.  Specifically, the models included in this figure are s12o[0-3], s20o[0-3], s40o[0-2], s60o[0-3], s12o[0,2,3]x, and s40o[0-1]$^{3D}$.  For the rapidly rotating 12 $M_\odot$ 2D SFHx runs, we notice very little EOS dependence on the normalized ramp-up slopes, as they nearly overlap with the corresponding SFHo runs.  While in general these slopes are driven by the cooling of the PNS---a process heavily dependent on the EOS---we notice differences in non-normalized slopes of only $\sim 3\%$ between SFHo and SFHx.

To explain the relationship seen in Figure \ref{fig:fdot_vs_J} and justify our mass cut of 1.75 $M_\odot$, we appeal to Figure \ref{fig:Jprofile}.  Displayed in the left panel of Figure \ref{fig:Jprofile} is the enclosed angular momentum ($J_\mathrm{enclosed}$) for the 60 $M_\odot$ progenitor, binned by a mass coordinate over 20 bins: specifically, models s60o0.5, s60o1, s60o2, and s60o3.  For increasing $\Omega_0$ we note a steeper $J_\mathrm{enclosed}$ profile.  In the right panel, we identically bin our data, but display $J_\mathrm{enclosed}$ for the four progenitor masses with $\Omega_0 = 1$ rad s$^{-1}$: specifically, models s12o1, s20o1, s40o1, and s60o1.  In this case, the higher the progenitor compactness, the steeper $J_\mathrm{enclosed}$ profile.  This behavior is expected because at collapse, the differential rotation parameter ($A$ in Equation (\ref{eq:omega})) is assigned based on initial compactness \citep{pajkos:2019}.  In general, with a larger differential rotation parameter (more solid body) we expect a steeper $J_\mathrm{enclosed}$ profile.  In both panels of Figure \ref{fig:Jprofile}, a significant amount of $J$ is deposited in the outer layers of the rotating supernova at the time of core bounce.  As mass is accreted, so too is $J$; this accretion will centrifugally support the contracting PNS.  The effect on the physical observable is a flatter $\dot{f}$.  With mass cuts less than $1.75\, M_\odot$, the tight linear relationship seen in Figure \ref{fig:fdot_vs_J} breaks down because less information about the accreted $J$ would be accounted for.  While a slightly higher mass cut could have been chosen, our computational domain only contains $\sim 1.85\,M_\odot$ of material, and this slight difference does not produce a noticeable change in Figure \ref{fig:fdot_vs_J}.  Moreover, depending on the progenitor, stellar material at that mass cut may not get accreted even hundreds of milliseconds after bounce.  Hence, we justify a mass cut of $1.75 \, M_\odot$.  Of course, for future work, optimizing this mass cut on computational domains that contain more mass could refine this analysis.

One large uncertainty that can affect $\dot{f}$ is the influence of angular momentum transport within stellar interiors \citep{aerts:2019}, so it is important to highlight how it may modify the relationship seen in Figure \ref{fig:fdot_vs_J}.  In our current simulation suite, $J$ is advected along with the fluid.  However, other influences that are not included in this work (such as magnetic fields) may change how $J$ is displaced throughout the supernova evolution.  In the event of stronger $J$ transport, less $J$ would be accreted onto the PNS.  As such, the PNS would receive less centrifugal support, allowing it to cool to smaller radii.  This effect would result in larger values of $\dot{f}$, even in the rapidly rotating cases.  Physically, this would diminish the flattening effect seen in rotating simulations.  Systematically, the points at large values of $J^\mathrm{bounce}_{1.75M_\odot}$ seen in Figure \ref{fig:fdot_vs_J} would shift upwards.

\subsection{Constraining the Stellar Core Mass Distribution}

\begin{table}[t]
\centering
\begin{tabular}{c|c|c|c}
Coefficient  & Value & Standard Deviation & Units\\
\hline
$\alpha$  &  $-1.17$ & $\pm 0.11$ & $10^{-168}(\mathrm{erg\, s})^{-3}$\\
$\beta$  &   $4.74$  & $\pm 0.27$ & $10^{-119}(\mathrm{erg\, s})^{-2}$\\
$\gamma$  &  $-0.315$ & $\pm 0.015$ & $10^{-49} \mathrm{(erg\, s)^{-1}}$  \\
$\delta $  &      925 & $\pm 87$ & Hz s$^{-1}$     \\
$\epsilon $  &    1760 & $\pm 50$ & Hz s$^{-1}$  
\end{tabular}
\caption{Coefficients used to constrain progenitor core compactness.}
\label{table:coefficients}
\end{table}

For clarity, we now outline the established three pieces of information from GWs emitted in a rotating CCSN.  We have shown that the amplitude of the bounce signal ($\Delta h$) correlates with $J_{1.75M_\odot}$ at bounce.  The ramp-up slope ($\dot{f}$) of the GW signal for nonrotating CCSNe relates to $\xi_{1.75}$.  And using our new findings, we quantify how much this slope is flattened, depending on $J_{1.75M_\odot}$.  We now synthesize these three points to place constraints on the mass distribution of the supernova progenitor.  Note in this section, unadorned $\beta$ refers to a polynomial fitting coefficient, whereas in previous sections, $\beta^\mathrm{bounce}_\mathrm{core}$ refers to the ratio of rotational kinetic energy to gravitational binding energy ($T/|W|$).

From examining the bounce signal, we have shown the relationship between core bounce and inner angular momentum as 
\begin{equation}
    \Delta h = \alpha J_{1.75M_\odot}^3 + \beta J_{1.75M_\odot}^2.
    \label{eq:h_J}
\end{equation}

From the accretion phase, we know the angular momentum quantifies how much the GW signal is flattened
\begin{equation}
    \dot{f}_J/\dot{f}_{J=0} = \gamma J_{1.75M_\odot} + 1.
    \label{eq:fdot_J}
\end{equation}

Lastly, we link core structure and nonrotating ramp-up slope
\begin{equation}
    \dot{f}_{J=0} = \delta \xi_{1.75} + \epsilon.
    \label{eq:fdot_xi}
\end{equation}

To begin, substitute Eqn. (\ref{eq:fdot_xi}) into Eqn. (\ref{eq:fdot_J}) and solve for $J_{1.75M_\odot}$ to yield
\begin{equation}
    J_{1.75M_\odot} = \frac{1}{\gamma}\Bigg(\frac{\dot{f}_J}{\delta \xi_{1.75} + \epsilon}  - 1\Bigg).
    \label{eq:J_xi}
\end{equation}

Next, substitute Eqn. (\ref{eq:J_xi}) into Eqn. (\ref{eq:h_J}) resulting in 
\begin{align}
    \Delta h = \alpha\mathcal{J}^3 + \beta\mathcal{J}^2,
    \label{eq:h_xi}
\end{align}

where $\mathcal{J}$ is represented by
\begin{equation*}
    \mathcal{J} = \frac{1}{\gamma} \Bigg(\frac{\dot{f}_J}{\delta \xi_{1.75} + \epsilon}  - 1\Bigg)
\end{equation*}

and the coefficients of interest are located in Table \ref{table:coefficients}.  Equation (\ref{eq:h_xi}) results in a cubic polynomial that has three unknowns: $\Delta h$, $\dot{f}_J$, and $\xi_{1.75}$.  In the event of a rotating CCSN detection, the two most distinct parts of the signal---$\Delta h$ and $\dot{f}_J$---can be directly obtained.  Thus, $\xi_{1.75}$ can be solved for using a numerical root finder.  

It is important to note that Equation (\ref{eq:fdot_J}), and the resulting analysis, solves for $\xi_{1.75}$ of a nonrotating progenitor.  To apply this solution to a hypothetical rotating CCSN, one needs to assume the core mass distributions are the same between a rotating and nonrotating case.  

In principle, this analysis solidifies how the fundamental rotational quantities $\Delta h$ and $\dot{f}$ relate to $\xi_{1.75}$.  In practice, however, we find the numerical inversion of Equation (\ref{eq:h_xi}) poses convergence challenges with numerical root finders in certain regions of parameter space because of its highly nonlinear nature.  In the next subsection, we describe how to streamline the analysis and remedy these convergence challenges.
\subsection{Estimating Stellar Properties Based on Compactness}

\begin{figure}
    \centering
    \includegraphics[width=0.5\textwidth,trim={20 20 0 10}]{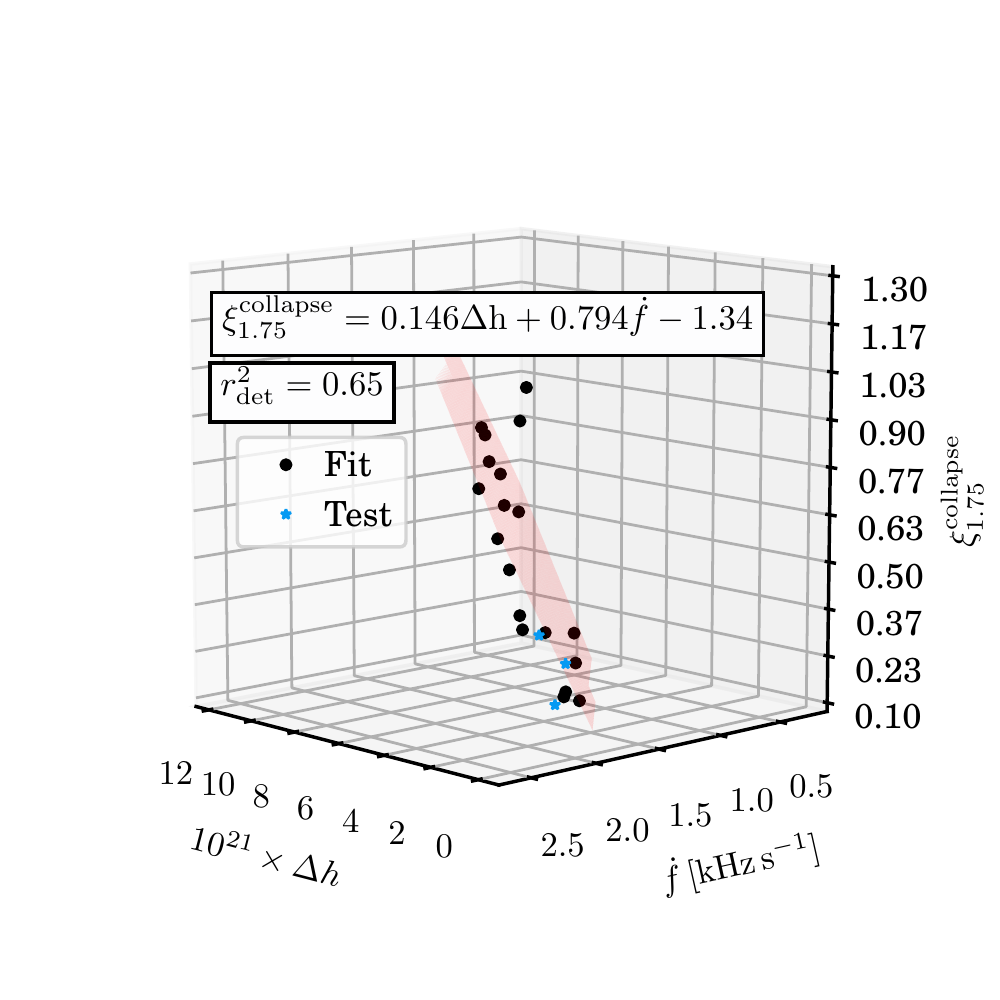} 
    \caption{Three-dimensional planar fit to the bounce amplitude, ramp-up slope, and core compactness at collapse.  Black dots denote data used to construct the planar fit (models s12o[0-3], s12o[0-3], s40o[0-2], and s60o[0-3]) or the fit data.  Cyan stars represent the test data, namely, the three 12 $M_\odot$ simulations using the SFHx EOS (models s12o[0,2,3]x) and are not involved in the fitting process, to reduce bias when verifying the planar fit.  The standard deviation errors of the fitting coefficients for the bounce, frequency, and vertical shift coefficients are $\sigma_h = \pm 0.026 $, $\sigma_f = \pm 0.150$, and $\sigma_s = \pm 0.35$, respectively.}
    \label{fig:3D_param}
\end{figure}

\begin{figure*}
    \centering
    \includegraphics{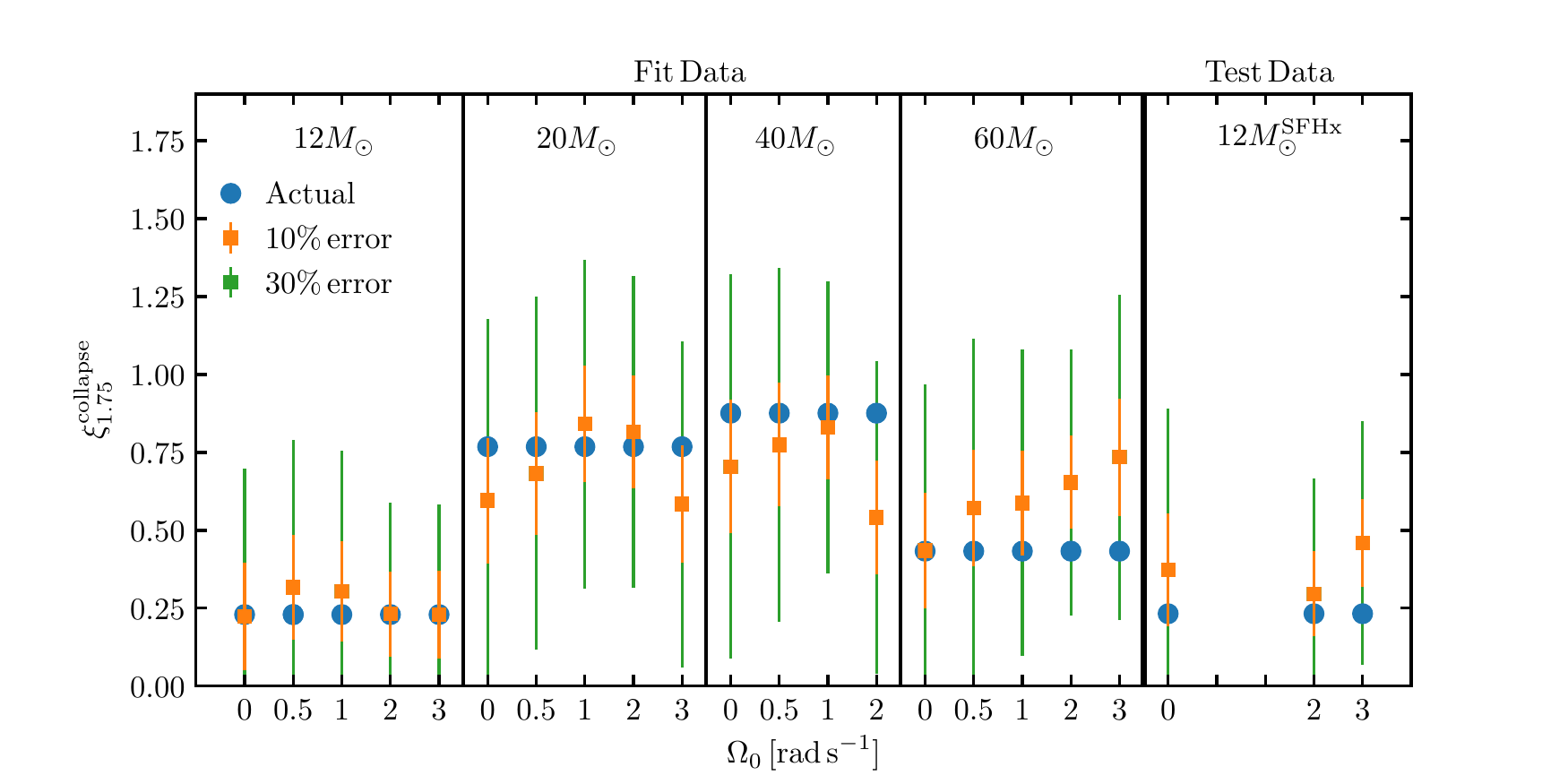}
    \caption{Input (actual) $\xi_{1.75}$ for progenitors at collapse compared against the estimated $\xi_{1.75}$ value using the planar fit, with identical $\Delta h$ and $\dot{f}$ inputs.  Orange (green) error bars represent one standard deviation of error assuming 10\% (30\%) uncertainty in the $\Delta h$ and $\dot{f}$ measurement.  Models included in the fit data are s12o[0-3], s12o[0-3], s40o[0-2], and s60o[0-3]. Models included in the test data are s12o[0,2,3]x. }
    \label{fig:xi_recon}
\end{figure*}

We begin inspecting the parameter space covered by $\Delta h$, $\dot{f}$, and $\xi_{1.75}$---visually represented in Figure \ref{fig:3D_param}.  In an attempt to prevent overfitting, while appropriately modeling the data, we construct a planar fit between these three variables (red plane in Figure \ref{fig:3D_param}) 

\begin{equation}
    \mathrm{\xi_{1.75}^{collapse}} = 0.146 \Delta h + 0.794 \dot{f} - 0.134 ,
    \label{eq:xi_plane}
\end{equation}
where $\Delta h$ is scaled by $10^{21}$ and $\dot{f}$ is in units of (kilohertz per seconds).  We treat our 19 2D SFHo models as the \textit{fit data} used to calculate the planar coefficients in Equation (\ref{eq:xi_plane}): models s12o[0-3], s12o[0-3], s40o[0-2], and s60o[0-3].  In an attempt to reduce bias from verifying our planar fit, we withhold 3 2D SFHx simulations (models s12o[0,2,3]x) from the fitting process, and reserve them as our \textit{test data} to test the reliability of Equation (\ref{eq:xi_plane}) for different EOSs.

Armed with Equation (\ref{eq:xi_plane}), we apply the previously used $\Delta h$ and $\dot{f}$ values to yield estimated $\xi_{1.75}$ values.  Figure \ref{fig:xi_recon} compares actual (blue dots) versus estimated (squares) $\xi_{1.75}$.  Included with the estimated values are the error bars assuming a standard deviation of $10\%$ ($30\%$) of both the $\Delta h$ and $\dot{f}$ measurements.  Convolved with the error of each planar coefficient yields error bars displayed in orange (green) that correspond to the standard deviation of the estimated $\xi_{1.75}$.  We note while the majority our test data and fit data are captured by the $10 \%$ error case, there are still outliers.  Returning to Figure \ref{fig:3D_param}, one can observe the presence of data points that deviate from the planar fit. While a more complex fit could mitigate the error, this raises the risk of overfitting.  Ideally, a larger number of simulations, spanning a finer resolution in parameter space would help estimate more accurate and statistically significant values for $\xi_{1.75}$.  Acknowledging areas for future improvement, we note the exciting implications this relationship has.
 
The mass distribution within a supernova progenitor is still highly uncertain and extremely difficult to constrain with EM observations alone.  The strength of this work is that it provides a framework to make the critical $\xi_M$ measurement solely using GWs.  This new analysis method now allows scientists to leverage various previous studies that have shown that $\xi_M$ has a significant impact on properties of the explosion \citep{oconnor:2011, sukhbold:2016}.  For example, \citet{sukhbold:2016} use a suite of simulations to connect $\xi_{2.5}$ to two physical observables: explosion energy and $^{56}$Ni yield.  Thus, as the needed GW signals for this analysis happen less then 0.5 s after core bounce, astronomers potentially can \textit{predict} future explosion properties of the supernova---helping provide valuable input for EM follow-up.  Furthermore, after the explosion has succeeded, making EM measurements of the explosion energy and $^{56}$Ni mass provide a testable case that can either validate or refine the accuracy of this GW prediction.

\subsection{Observability of GW Signal}


When discussing possible GW signals from rotating CCSNe, it is important to anticipate the likelihood of observation.  Current GW detectors are limited to Galactic core-collapse events--an estimated rate of $\sim2$ per century \citep{diehl:2006}.  Convolved with the estimate that only 1\% of massive stars can reach the rapid rotation regime, this estimate drops to a rapidly rotating, Galactic CCSN rate of 2 every 10,000 yr.  Nevertheless, due to the poorly understood influence of binarity and magnetic braking, this estimate could be higher \citep{woosley:2006, demink:2013}.  Likewise, the possible GW signals from so-called \textit{failed supernovae} also potentially can act as sources of detectable GWs \citep{fryer:2003}.  
Another key consideration are the effects of viewing angle.  As outlined in \citet{oohara:1997}, the relative orientation between the CCSN and observer impacts the amplitude of the GW signal measured.  As such, the $\dot{f}$ measurement should be unaffected.  By contrast, the measured $\Delta h$ for this analysis is completely degenerate with detector orientation.  In all likelihood, the equator of an arbitrarily oriented CCSN will be off axis with GW detectors.  This viewing angle effect would yield a smaller amplitude detected for $\Delta h$.  Hence, in the conservative case where no orientation information can be gathered, a given $\Delta h$ would serve as a lower limit for the $\xi_{1.75}$ estimate.   

An alternative to help constrain CCSN orientation is the use of neutrinos.  For a rotating CCSN,  depending on the degree of differential rotation, the shape of the neutrinosphere can become deformed, namely, oblate \citep{kotake:2003b}.  The resulting nonspherical emission of neutrinos in principle contains information about the supernova orientation.  More recently, \citet{nagakura:2021} more concretely noted the angular variations in the event rate and---more modestly---in the time-integrated signal.  While research into the dependence of neutrino emission based on viewing angle is still ongoing, its continued progress and recent advances stand as a promising sign to one day help constrain CCSN properties purely using multimessenger methods. 

In the rotating core-collapse scenario, there are two main sources of detectable GWs that occur at different times.  The first is the core-bounce signal, which occurs immediately after core bounce.  The second is during the accretion phase of the supernova.  As seen in Figure \ref{fig:spectrogram}, these fundamental modes of the PNS can last hundreds of milliseconds and display an increase in frequency as the PNS cools.  Likewise, in some cases, rotational instabilities can induce GW production with similar amplitudes to the bounce signal, when viewed along the axis of rotation \citep{scheidegger:2010,kuroda:2014}.

In order for a GW observatory to successfully \textit{detect} a GW event, two factors must be considered: detector sensitivity and signal reconstruction \citep{abbott:2016b}.  Detector sensitivity limits the frequency range of potential GW signals as well as GW strengths, due to signal-to-noise (S/N) constraints.  Signal reconstruction is the act of separating the GW signal from the detector noise and is highly sensitive to S/N constraints as well \citep{mciver:2015}.  Thus, even though a given CCSN may emit multiple GW modes at a variety of frequencies, it is likely only the dominant features of the GW signal will be reconstructed.

These observational considerations lie at the heart of this project.  Whereas other works are limited to progenitors of a certain compactness or depend on observing multiple oscillatory modes of the PNS, ours depends only on the dominant bounce signal as well as the main ramp-up of the PNS.  In extreme cases, although rare, the amplitude of the bounce signal from rotating CCSNe can be nearly an order of magnitude larger than the GWs from PNS oscillations, as seen in Figure \ref{fig:TDWF_all}.  This effect can push the detection volume out by a corresponding order of magnitude because the GW amplitude scales as the inverse of the distance to the source. 

\section{Summary and Conclusion}
\label{sec:summary}

We present a new method to constrain supernova progenitor compactness at collapse by using information from two parts of the GW signal: the core bounce and slope of the early ramp-up.  Our findings are summarized as follows.  

\begin{itemize}
    \item We highlight the importance of robust treatments of deleptonization (eg., M1) on collapse, as it impacts the magnitude and directional dependence of the $Y_e(\rho)$ profile of the CCSN core and consequently $\Delta h_\mathrm{bounce}$, seen in Figure \ref{fig:ye_vs_rho}.

    \item We build upon the findings of \citet{warren:2020} by corroborating that the slope of the early ramp-up of nonrotating CCSNe correlates with compactness of the inner $1.75 \,M_\odot$ in 2D simulations, seen in Figure \ref{fig:fdot_vs_xi}.
    
    \item We relate the amplitude of the bounce signal to the angular momentum of the inner $1.75 \,M_\odot$, seen in Figure \ref{fig:hb_vs_J}.
    
    \item We quantify the dependence of PNS ramp-up slope on the angular momentum of the inner $1.75 \,M_\odot$ at bounce, seen in Figure \ref{fig:fdot_vs_J}.
    
    \item We combine the two parts of a given rotating CCSN GW signal---bounce and main ramp-up slope---to constrain progenitor core structure by estimating $\xi_{1.75}$, seen in Figure \ref{fig:3D_param}.
    
    \item For a rotating CCSN, these two parts of the GW signal are the most likely to be detected and reconstructed by current GW detectors.
    
    \item Because the GW signal used is emitted $\lesssim 0.5$ s after core bounce, we provide astronomers predictive power for EM emission, leveraging other works making EM correlations with $\xi_M$. 
    
\end{itemize}

While we have introduced a new method that uses GWs to constrain supernova properties and potentially predict their behavior, it is important to outline the limitations of this work.  The majority of this work was completed using axisymmetric simulations.  While previous works have used axisymmetry to predict bounce signals for a variety of rotational configurations \citep{abdik:2014}, 3D simulations remain the gold standard.  Later in the supernova, for the GW signal during the accretion phase, certain instabilities may arise---low $T/|W|$ or spiral SASI---that are inherently three dimensional.  Nevertheless, including four 3D models does corroborate the relationships between $J$, $\dot{f}$, and $\Delta h$ noted in this work.  We also do not include magnetic fields.  
Two instabilities that may arise when including magnetic fields are the $\alpha-\Omega$ dynamo \citep{mosta:2015} and MRI \citep{akiyama:2003}.  
While the impact of the $\alpha-\Omega$ dynamo and the MRI on dynamically relevant timescales is uncertain, these instabilities may drive convection in the post-shock region shortly after core bounce \citep{bonanno:2005,cerda-duran:2007}.    
Our simulations also use the GREP, neglecting full GR.  
While the GREP could affect $\dot{f}$ by overestimating the peak GW frequency when compared to GR, it has been shown to produce GW amplitudes of similar scale and similar PNS compactness \citep{muller:2013}.  In principle, this could affect the correlations presented in this work.  Likewise, exploring a wider range of EOSs could help isolate the slight EOS dependence of $\dot{f}$.

This work also provides scientific motivation to continue research into electron capture rates onto heavy nuclei.  As noted in Section \ref{subsec:bounce}, we observed differences in $\Delta h$ at bounce up to $40\%$ for identical rotation profiles with differing electron capture rates.  As $\Delta h$ at bounce is a fundamental parameter in determining $\xi_{1.75}$, more accurate rate measurements will allow astronomers to better constrain progenitor core compactness for a future Galactic CCSN.

We emphasize that this paper takes an empirical approach to finding these correlations.  We do not rely on assuming functional forms for quantities such as the moment of inertia at the center of the supernova.   This feature is advantageous for future GW observations of rotating CCSNe because it allows us to probe the mass distribution within a progenitor without deciding a priori how the mass may be distributed.  

We acknowledge the robustness of these fits can be increased by including more 3D simulations with higher fidelity treatments of gravity and magnetic fields.  Nevertheless, the focus of this work is to show that the accretion phase signal indeed contains information about the structure of CCSN progenitors.  By quantifying this information and combining it with rotational information encoded in the bounce signal, astronomers take one step closer toward determining the physical conditions that set the stage for the onset of stellar explosions.

\acknowledgements

  We thank the anonymous referee for their constructive feedback.  We thank Sheldon Wasik for helping us access the modified electron capture rates and Brandon Barker for assisting with the statistical analysis.  M.A.P. was supported by a Michigan State University Distinguished Fellowship.  M.L.W. is supported by an NSF Astronomy and Astrophysics Postdoctoral Fellowship under award AST-1801844.
  S.M.C. is supported by the U.S. Department of Energy, Office of Science, Office of Nuclear Physics,      Early Career Research Program under Award Number DE-SC0015904. This material is based upon work       supported by the U.S. Department of Energy, Office of Science, Office of Advanced Scientific          Computing Research and Office of Nuclear Physics, Scientific Discovery through Advanced Computing     (SciDAC) program under Award Number DE- SC0017955. 
This research was supported by the Exascale Computing Project (17-SC-20-SC), a collaborative effort of two U.S. Department of Energy organizations (Office of Science and the National Nuclear Security Administration) that are responsible for the planning and preparation of a capable exascale ecosystem, including software, applications, hardware, advanced system engineering, and early testbed platforms, in support of the nation's exascale computing imperative.  E.O.C. is supported by the Swedish Research Council (Project No. 2018-04575).  K.C.P. is supported by the Ministry of Science and Technology of Taiwan through grants MOST 107-2112-M-007-032-MY3.
The software used in this
work was in part developed by the DOE NNSA-ASC OASCR Flash Center at
the University of Chicago.  

    \software{FLASH (see footnote 7) \citep{fryxell:2000,fryxell:2010}, Matplotlib\footnote[8]{\url{https://matplotlib.org/}} \citep{hunter:2007},
    NuLib\footnote[9]{\url{http://www.nulib.org}}
    \citep{oconnor:2015},
    NumPy\footnote[10]{\url{http://www.numpy.org/}} \citep{vanderwalt:2011}, SciPy\footnote[11]{\url{https://www.scipy.org/}} \citep{jones:2001}}


\bibliography{ms}



\end{document}